\documentclass[aps,prd,twocolumn,nofootinbib,floatfix]{revtex4-2}
\pdfoutput=1
\usepackage{graphicx}  
\usepackage{dcolumn}  
\usepackage{bm}        
\usepackage{amsmath}   
\usepackage{amssymb}   
\usepackage{hyperref} 
\usepackage{color}     
\usepackage{geometry}
\usepackage{tikz}
\usepackage[compat=1.1.0]{tikz-feynman}
\usepackage{booktabs}
\usepackage{amsmath}   
\usepackage{slashed}  
\usepackage{orcidlink}
\allowdisplaybreaks

\tikzset{
	operator/.style={
		rectangle,
		fill=black,
		inner sep=2.5pt,
		anchor=center
	}
}

\hypersetup{
	colorlinks=true,
	linkcolor=blue,
	citecolor=blue,
	urlcolor=blue
}

\begin{document}
		
	\title{Branching ratios and CP violations of $B \to\rho (\omega) \gamma$ decays in the modified perturbative QCD approach and the relevant dark photonic decays of $B\to \rho (\omega) \gamma^\prime $ }
	
	\author{Shang Qi}\email{qis@mail.nankai.edu.cn}
	%\affiliation{School of Physics, Nankai University, Tianjin 300071, People's Republic of China}
	\author{Mao-Zhi Yang,\orcidlink{0000-0002-5052-713X}} \email{yangmz@nankai.edu.cn}
	\affiliation{School of Physics, Nankai University, Tianjin 300071, P. R. China}

	\date{\today}

	\begin{abstract}
In this work we study the radiative decays of $B\to \rho (\omega)\gamma$ processes in the modified perturbative QCD approach, where the transverse momenta of the quark and gluons envolved in the interaction process are kept, and the Sudakov factor is incorporated in the theoretical calculation, which are helpful to suppress the infrared end-point contribution. A critical infrared momentum cutoff scale is introduced, which is used to separate the soft and hard contributions in QCD. Then the form factors envolved in the radiative decays should be separated as hard and soft form factors. The hard form factors can be calculated with the perturbative QCD approach, while the soft form factors should be viewed as soft input parameters. The soft form factors can be obtained by confronting the theoretical calculation of the branching ratios of the radiative decay of $B$ meson  with the experimental data. Summing the soft and hard form factors, we find that the total form factors are in fine agreement with that obtained by nonperturbative theoretical method, such as the light-cone sum rules of QCD. We also study $B\to \rho (\omega) \gamma^\prime$ decays by using the form factors obtained in the $B\to \rho (\omega)\gamma$ decay processes, where $\gamma^\prime$ is the dark photon introduced by the extra $U(1)$ symmetry model. The upper limit of the branching ratios of thes decays are estimated. 
	\end{abstract}

	\maketitle
	
\section{Introduction}
\label{sec:intro}
Radiative decays $B\to \rho(\omega)\gamma$ are processes induced by the flavor-changing-neutral-current (FCNC) transition of $b\to d\gamma$ in the quark level, which is an interesting process for investigations of flavor physics. On one hand, it is an excellent platform to test the electroweak interaction and QCD effect in the standard model (SM); on the other hand, it is one of the most appropriate process to detect new physics because it is a pure loop effect which is sensitive to physics beyond the SM \cite{hurth2003}.  The FCNC process has attracted large amount of interest in theory. It has been considered up to next-to-leading order in QCD within the frame work of QCD factorization (QCDF) \cite{Bosch-Buchalla2002,Ali-etal2004}, and in effective theory of QCD, such as the large-energy-effective theory \cite{Ali-etal2002}. QCD factorization is a theoretical approach based on the factorization theorem with respect to the collinear part of the parton momentum \cite{Beneke-1999br,Beneke-2000ry,Beneke-2001ev,Beneke-2003zv}. Some dominant long-distance contributions beyond QCDF are considered in Ref. \cite{Ball-etal2007}. There are also studies of $B\to (K^*,\rho,\omega)\gamma$ decays with perturbative QCD (PQCD) method in Refs. \cite{Keum-etal2005,Lu-etal2005,Matsu-2006}, which is based on $K_\mathrm{T}$ factorization, where the transverse momenta of the partons are included in the calculation of the decay amplitude \cite{Li-Yu1995,Li1995,Li-Yu1996}.

The branching ratios and $CP$ violations for these radiative decays of $B$ meson have been measured in experiments. The world averaged data are \cite{PDG2024}
\begin{eqnarray}
Br(B^{+}\to\rho^{+}\gamma)&=(1.29\pm 0.20)\times10^{-6} \nonumber\\ 
Br(B^{0}\to\rho^{0}\gamma)&=(0.82\pm 0.13)\times10^{-6} \\
Br(B^{0}\to\omega\gamma)&=(0.44\pm 0.17)\times10^{-6} \nonumber
\end{eqnarray}	
for the branching ratios, and
\begin{equation}
A_{CP}(B^{+}\to\rho^{+}\gamma)=(-0.08\pm 0.15) 
\end{equation}	
for the $CP$ violation.

The branching ratios of $B\to (\rho,\omega)\gamma$ predicted by QCDF can approximately agree with the experimental data \cite{Bosch-Buchalla2002,Ali-etal2004,Ball-etal2007}. However, the predictions to the branching ratios of $B\to (\rho,\omega)\gamma$ by PQCD approach in Ref. \cite{Lu-etal2005} suffer large uncertainty, and the central values are too large compared with the experimental data. Therefore, it is necessary to improve the calculations of the branching ratios and $CP$ violations for these radiative decays in PQCD approach. 

$B\to\pi$ transition from factor has been analyzed in PQCD approach with several modifications in Ref. \cite{Lu-Yang2021}, where the wave function for $B$ meson is changed to be the one that is obtained by solving the wave equation in QCD-inspired relativistic potential model \cite{Yang2012,LY2014,LY2015,SY2017}, and the transition form factor is separated as hard and soft parts. The hard part of the form factor can be calculated with perturbative method, where the strong coupling constant is small $\alpha_s/\pi<0.2$. On the other hand, the soft part is controlled by nonpertubative dynamics. This is equivalent to introduce a critical infrared momentum cutoff scale $\mu_c$. The hard form factor is controlled by interactions with the scale $\mu>\mu_c$, and the soft form factor is relevant to interactions of lower scale $\mu<\mu_c$. The cutoff scale can be approximately taken as 1 GeV in practice.

In this work, we revisit the radiative decays of $B\to\rho (\omega)\gamma$ in PQCD approach with the above modifications. The $B$ meson wave function is the one that is obtained from the QCD-inspired relativistic potential model \cite{Yang2012,LY2014,LY2015,SY2017}, and an infrared momentum cutoff scale $\mu_c$ is introduced to separate the hard and soft interactions. The contributions to the form factors with the scale larger than $\mu_c$ are calculated by perturbative method, while the contributions with the scale lower than $\mu_c$ is replaced by the soft form factor, which is viewed as nonperturbative input.

The remainder of the paper is organized as follows. Section II is for the calculation in perurbative QCD in leading order (LO). Section III is devoted to the calculations of the next-to-leading order (NLO) in QCD. The annihilation contributions are considered in Sec. IV. Section V is for the contribution of soft form factor. Section VI is for the numerical treatment and discussion. The radiative decays of $B$ meson to dark photon are presented in Sec. VII by using form factors obtained in Sec. VI.  Finally section VIII is a brief summary.

\section{THE HARD AMPLITUDE OF LEADING-ORDER IN PERTURBATIVE QCD}
	\label{sec-hard}
	
	\subsection{The effective Hamiltonian}
The effective Hamiltonian describing $b\to d$ transition process is \cite{Buchalla1995}
\begin{equation} \label{Heff}
	\begin{split}
		&H_{\text{eff}} = \frac{G_F}{\sqrt{2}} \Bigg[ 
		V_{ub} V_{ud}^* \left( C_1^{(u)} O_1^{(u)} + C_2^{(u)} O_2^{(u)} \right) \\
		&\quad + V_{cb} V_{cd}^* \left( C_1^{(c)} O_1^{(c)} + C_2^{(c)} O_2^{(c)} \right) - V_{tb} V_{td}^*  \\
		&\quad\cdot\left( \sum_{i=3}^{6} C_i O_i + C_{7\gamma} O_{7\gamma}+ C_{8g} O_{8g} \right) 
		\Bigg]
	\end{split}
\end{equation}
where $G_F$ is the Fermi constant, $V_{qb} V_{qd}^* $ ($q=u,c,t$) the product of the Cabibbo-Kobayashi-Maskawa (CKM) matrix elements, $C_i$'s the Wilson coefficients, and $Q_i$'s the effective operators, which are
\begin{equation}
	\begin{aligned}
		O_1^{(q)} &= (\bar{d}_i q_j)_{V-A} (\bar{q}_j b_i)_{V-A}, \\
		O_2^{(q)} &= (\bar{d}_i q_i)_{V-A} (\bar{q}_j b_j)_{V-A}, \\
		O_3^{(q)} &= (\bar{d}_i b_i)_{V-A} \sum_{q} (\bar{q}_j q_j)_{V-A}, \\
		O_4^{(q)} &= (\bar{d}_i b_j)_{V-A} \sum_{q} (\bar{q}_j q_i)_{V-A}, \\
		O_5^{(q)} &= (\bar{d}_i b_i)_{V-A} \sum_{q} (\bar{q}_j q_j)_{V+A}, \\
		O_6^{(q)} &= (\bar{d}_i b_j)_{V-A} \sum_{q} (\bar{q}_j q_i)_{V+A}, \\
		O_{7\gamma} &= \frac{e}{8\pi^2} m_b \bar{d}_i \sigma^{\mu\nu} (1 + \gamma_5) b_i F_{\mu\nu}, \\
		O_{8g} &= \frac{g}{8\pi^2} m_b \bar{d}_i \sigma^{\mu\nu} (1 + \gamma_5) T^a_{ij} b_j G^a_{\mu\nu},
	\end{aligned}
	\label{eq:operators} % 给这个整体加一个标签，方便文中引用
\end{equation}
with $i,j$ being the color indices, $e$ the electromagnetic coupling, $g$ the strong coupling, and $F_{\mu\nu}$ and $G^a_{\mu\nu}$ the strength tensors of the electromagnetic and color fields, respectively. And the current $(\bar{q}_i q_j)_{V\pm A}$ denotes $\bar{q}_i \gamma^\mu (1\pm \gamma_5) q_j$.

With the effective Hamiltonian given in Eq. (\ref{Heff}), the decay amplitude of the $B\to \rho (\omega)\gamma$ decay can be written as
\begin{equation}
\begin{split}
	A =  &\langle F | H_{\text{eff}} | B \rangle= \frac{G_F}{\sqrt{2}} \sum_{i,q} V_{qb}^* V_{qd} C_i(\mu)\\ &\cdot\langle F | O_i(\mu) | B \rangle,
\end{split}
\end{equation}
where $F$ stands for the final state $\rho\gamma$ and $\omega\gamma$ in the decays.

The decay amplitude $A$ can be decomposed into the following from according to its Lorentz structure
\begin{equation}\label{MSMP}
	A = (\varepsilon_V^* \cdot \varepsilon_\gamma^*) M^S + \frac{i}{P_V \cdot P_\gamma} \epsilon_{\mu\nu\rho\sigma} \varepsilon_\gamma^{*\mu} \varepsilon_V^{*\nu} P_\gamma^\rho P_V^\sigma M^P,
\end{equation}
where $P_V$ is the momentum of the vector meson $\rho$ and/or $\omega$, $P_\gamma$ the momentum of the photon, $\varepsilon_V$ and $\varepsilon_\gamma$ the polarization vectors of the vector meson and photon, respectively.

The matrix element $\langle F | H_{\text{eff}} | B \rangle$ can be calculated in PQCD approach for the contributions from the interactions transferred by hard gluon. It is convenient to use the light-cone coordinate in the calculations. Then any momentum $p$ can be expressed in the form
\begin{equation}
	p = (p^+, p^-, \vec{p}_T) = \left( \frac{p^0 + p^3}{\sqrt{2}}, \frac{p^0 - p^3}{\sqrt{2}}, (p^1, p^2) \right),
\end{equation}
and the product of any two momenta $A$ and $B$ can be expressed in terms of the light-cone components equivalently as $A\cdot B= A_\mu B^\mu=A^+B^-+A^-B^+ -\vec{A}_\perp\cdot \vec{B}_\perp$.

In the $B$ meson rest frame, the momenta of $B$, $\rho$ ($\omega$) and the photon $\gamma$ can be written as
\begin{equation}
\begin{split}
&P_B=(P^+_B,P^-_B,\vec{P}_{B\perp})=\frac{M_B}{\sqrt{2}}(1,1,\vec{0}_\perp),\\
&P_V=(P^+_V,P^-_V,\vec{P}_{V\perp})=\frac{M_B}{\sqrt{2}}(0,1,\vec{0}_\perp),\\
&P_\gamma=(P^+_\gamma,P^-_\gamma,\vec{P}_{\gamma\perp})=\frac{M_B}{\sqrt{2}}(1,0,\vec{0}_\perp),
\end{split}
\end{equation}
where the vector mesons $\rho$ or $\omega$ are chosen as moving in the ``$-$ " direction of the $z$-axis and the photon in the ``$+$" direction. The momenta of the spectator quarks in $B$ and the vector mesons are
\begin{equation}
\begin{split}
&k_1=(k^+_1,k^-_1,\vec{k}_{1\mathrm{T}})=(\frac{M_B}{\sqrt{2}}x_1,0,\vec{k}_{1\mathrm{T}}),\\
&k_2=(k^+_2,k^-_2,\vec{k}_{2\mathrm{T}})=(0,\frac{M_B}{\sqrt{2}}x_2,\vec{k}_{2\mathrm{T}}),
\end{split}
\end{equation}
where $x_1$ and $x_2$ are the momentum fractions defined by $x_1=k_1^+/P^+_B$ and $x_2=k_2^-/P^-_V$, respectively.

\subsection{The factorization formula for the decay amplitude
		and the meson wave functions}

The mass of $b$ quark is significantly larger than that of light quarks $u$, $d$ and $s$, so the energy release in the decays of $B$ meson to light mesons is usually large. Then the momenta transferred by the gluons between the spectator quark and the other quarks (or antiquarks) in $B$ or the final mesons are also generally large to kick the light quark in the rest $B$ meson to a speed which matches the fast moving light final particles. Therefore, hard interaction generally dominates in $B$ decays, which can be treated by perturbative QCD. The PQCD factorization scheme has been developed for $B$ decays several decades before \cite{Li-Yu1995,Li1995,Li-Yu1996}. Soft gluon exchanges between the quark lines can generate the large double logarithms like $\ln^2(Pb)$ as the collinear and soft divergences overlap, with $P$ being the dominant light-cone component of a meson momentum, and $b$ the conjugate variable of the transverse momentum $k_\mathrm{T}$ of the quark or antiquark in a meson.The resummation of these double logarithms results in the Sudakov factor $\exp^{[-s(P,b)]}$. The Sudakov factor can suppress the long-distance contributions as $b$ being large, and vanishes when $b>1/\Lambda_\mathrm{QCD}$. Loop corrections can also result in large double logarithms as $\alpha_s \ln^2 x$ in the end-point region $x\to 0$, where $x$ is the momentum fraction of the quark or antiquark in a meson, and the summation of these logarithms leads to the threshold factor $S_t(x)$ \cite{Li2002}. Both the Sudakov and threshold factors can help to suppress the long-distance contributions and the explicit form of them can be found in Appendix \ref{Appendix A}.

The factorization formula for $B\to\rho(\omega)\gamma $ decay can be written in $b$-space as
\begin{equation}\label{eq-factorization}
	\begin{split}
 &\mathcal{M}= \int_{0}^{1} dx_1 dx_2 \int_{0}^{1/\Lambda} d^2b_1 d^2b_2  C(\mu) \\
		&\quad \times\exp^{-S(x_1,x_2,b_1,b_2,\mu)}  \Phi^{V}(x_2, b_2) \\
           &\quad\times H(x_1, x_2, b_1, b_2, \mu)  \Phi^{B}(x_1, b_1),
	\end{split}
\end{equation} %\otimes
where $C(\mu)$ stands for a collect of the relevant Wilson coefficients, and $\exp^{-S(x_1,x_2,b_1,b_2,\mu)}$ the Sudakov factor. $\Phi^{V}(x_2, b_2)$ and $\Phi^{B}(x_1, b_1)$ are the wave functions of the vector and $B$ mesons in the conjugate $b$-space, respectively. $H(x_1, x_2, b_1, b_2, \mu) $ is the hard kernel from the hard scattering contribution. $b_1$ and $b_2$ are the conjugate variables to the transverse momenta of the spectator quarks in $B$ and vector mesons, and $\Lambda=\Lambda_\mathrm{QCD}$.

The wave function of $B$ meson in the momentum space can be define by the matrix element
\begin{equation}
	\langle 0| \bar{q}(z)_\beta [z, 0] b(0)_\alpha | \bar{B} \rangle = \int d^3 k \Phi_{\alpha\beta}^B (\vec{k}, \mu) e^{-i k \cdot z},
\end{equation}
where $\alpha$ and $\beta$ are the spinor indices, and $\Phi_{\alpha\beta}^B (\vec{k})$ being the spinor wave function of $B$ meson, which is given by 
\begin{equation}
	\begin{split}
		\Phi^{B}_{\alpha\beta}(\vec{k}) &= \frac{-i f_B m_B}{4} K(\vec{k}) \\
		&\quad \cdot \Bigg\{ (E_Q + m_Q) \frac{1+\not{v}}{2} \left[ \left( \frac{k^+}{\sqrt{2}} + \frac{m_q}{2} \right) \not{n}_+ \right. \\
		&\quad \left. + \left( \frac{k^-}{\sqrt{2}} + \frac{m_q}{2} \right) \not{n}_- - k_\perp^\mu \gamma_\mu \right] \gamma^5 \\
		&\quad - (E_q + m_q) \frac{1-\not{v}}{2} \left[ \left( \frac{k^+}{\sqrt{2}} - \frac{m_q}{2} \right) \not{n}_+ \right. \\
		&\quad \left. + \left( \frac{k^-}{\sqrt{2}} - \frac{m_q}{2} \right) \not{n}_- - k_\perp^\mu \gamma_\mu \right] \gamma^5 \Bigg\}_{\alpha\beta},
	\end{split}
\end{equation}
where $f_B$ is the decay constant of $B$ meson, $m_B$ the $B$ meson mass, $E_Q$ and $m_Q$ the energy and mass of the heavy quark $b$, respectively. $k$ is the momentum of the light quark in the $B$ meson. $E_q$ and $m_q$ are the energy and mass of the light quark in $B$ meson, respectively. The light quark mass $m_q$ is neglected in the numerical calculation. $v$ is the four-speed of the $B$ meson, such that $p_B=m_B v$. Two light-like vectors are defined as $n_{\pm}^\mu = (1, 0, 0, \mp 1)$, and $k^\pm = \frac{E_q \pm k^3}{\sqrt{2}}$, $	k_\perp^\mu = (0, k^1, k^2, 0)$.

The function $K(\vec{k})$ is
\begin{equation}
	K(\vec{k}) = \frac{2 N_B \Psi_0(\vec{k})}{\sqrt{E_q E_Q (E_q + m_q)(E_Q + m_Q)}},
\end{equation}
where $N_B=\frac{i}{f_B}\sqrt{\frac{3}{(2\pi)^3 m_B}}$, and $\Psi_0(\vec{k})$ is the wave function for $B$ meson, which has been obtained by numerically solving the wave equation in the relativistic potential model \cite{Yang2012,LY2014,LY2015}. An analytical form for this wave function has been obtained in Refs. \cite{SY2017,SY2019} by fitting to the numerical solution in the potential model, which is
\begin{equation}
	\Psi_0(\vec{k}) = a_1 e^{a_2 |\vec{k}|^2 + a_3 |\vec{k}| + a_4}
\end{equation}
where the parameters are \cite{SY2017}
\begin{equation}
\begin{aligned}
	a_1 &= 4.55^{+0.40}_{-0.30} \, \text{GeV}^{-3/2}, &
	a_2 &= -0.39^{+0.15}_{-0.20} \, \text{GeV}^{-2}, \\
	a_3 &= -1.55 \pm 0.20 \, \text{GeV}^{-1}, &
	a_4 &= -1.10^{+0.10}_{-0.05}.
\end{aligned}
\end{equation}

In the processes of $B\to\rho(\omega)\gamma$ decays, the vector mesons $\rho$ and $\omega$ are transversely polarized. The spinor wave function of vector meson in transverse polarization is defined as  \cite{Ball-1998,KLS-2001}
%\begin{widetext}
	\begin{equation}
     \begin{split}
	&	\langle \rho/\omega(P, \epsilon_V^T) | \bar{d}_\alpha(z) u_\beta(0) | 0 \rangle\\
 &= \frac{1}{\sqrt{2N_c}} \int_0^1 dx e^{ixP \cdot z} 
		\Bigg[ M_V \not{\epsilon}_V^{*T} \phi_V^v(x)   \\
		&+ \not{\epsilon}_V^{*T}\not{P} \phi_V^T(x)- \frac{M_V}{P \cdot n_+} i \epsilon_{\mu\nu\rho\sigma} \gamma^5 \gamma^\mu  \\
      &\cdot \epsilon_V^{T\nu} P^\rho n_+^\sigma \phi_V^a(x) \Bigg],
      \end{split}
	\end{equation}
%\end{widetext}
where $\epsilon_V^{T}$ is the polarization vector of the vector meson in transverse polarization. $\phi_V^v(x)$ is the distribution amplitude of twist-2, and $\phi_V^T(x)$ and $\phi_V^a(x)$ are the distribution amplitude of twist-3. Their expressions in terms of Gegenbauer polynomials can be found in Appendix \ref{Appendix B}. 

\subsection{The leading order contribution}

In this section we present the decay amplitudes of the leading order contributions in QCD, which are contributed by the operators of $O_{7\gamma}$ and $O_{8g}$. 
\begin{figure}[htbp]
	\centering
	% 直接插入刚才生成的 PDF 文件
	\includegraphics[width=7.5cm]{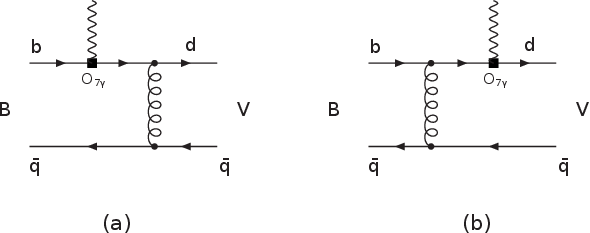}
	\caption{Contribution of operator $O_{7\gamma}$ to
		$B\to\rho (\omega )\gamma $ decay.}\label{fig-O7gamma}
	%\label{fig:feynman1}
\end{figure}

The diagrams of the contribution of $O_{7\gamma}$ are depicted in Fig.\ref{fig-O7gamma}, where hard gluon is exchanged between the spectator and other quarks. The contributions of the operator $O_{7\gamma}$ to the scalar and pseudoscalar parts of the amplitude $M^S$ and $M^P$ defined in Eq. (\ref{MSMP}) are 

%\begin{widetext}
	\begin{equation}\label{M7gamma-a}
		\begin{aligned}
&\mathcal{M}^{S(a)}_{7\gamma} = -\frac{2 e m_B^4 f_B G_F m_0}{\sqrt{27}} \eta_t \int_0^\infty k_T dk_T \int_{x^d_1}^{x^u_1}dx_1  \\
&\times \int^1_0 dx_2 \int^{1/\Lambda}_0 b_1 db_1b_2 db_2 \, \alpha_s(t^a_7) \, C_{7\gamma}(t^a_7) \left( \frac{m_B}{2} \right. \\
&+\left. \frac{k_T^2}{2 x_1^2 m_B} \right)((E_q - k_z)E_Q + m_Q) \left[ \phi_V^a(x_2, b_2)\right.   \\
& +\left. \phi_V^v(x_2, b_2) \right]	K(k) \, J_0(k_T b_1)\, H^a_7(x_1, x_2, b_1, b_2) \\
&\times\exp\left[ -S_B(t^a_7) - S_V(t^a_7) \right] \, S_t(x_1),
		\end{aligned}%\nonumber
	\end{equation}
	
\begin{equation}\label{M7gamma-b}
	\begin{aligned}
	&\mathcal{M}^{S(b)}_{7\gamma} = -\frac{2 e m_B^4 f_B G_F}{\sqrt{27}} \eta_t \int_0^\infty k_T dk_T\int^{x^u_1}_{x^d_1} dx_1  \\
	&\times \int_0^1dx_2 \int_0^{1/\Lambda} b_1 db_1 b_2 db_2  \, \alpha_s(t^b_7) \, C_{7\gamma}(t^b_7) \left( \frac{m_B}{2}\right. \\
&+\left. \frac{k_T^2}{2 x_1^2 m_B} \right) (E_Q + m_Q) \bigg\{ m_B \left[ E_q(1+x_2)\right. \\
\end{aligned}\nonumber
\end{equation}
\begin{equation}
\begin{aligned}
&+ \left. k_z(1-x_2) \right] \phi_ t(x_2, b_2) - m_0 \left[ E_q(2x_2-1)\right.  \\
& +\left. k_z \right] \left[ \phi_V^a(x_2, b_2) + \phi_V^v(x_2, b_2) \right] \bigg\}K(k) \, J_0(k_T b_1) \, \\
& \times  H^b_7(x_1, x_2, b_1, b_2) \exp\left[ -S_B(t^b_7) - S_V(t^b_7) \right] \\
&\times  S_t(x_2),
	\end{aligned}
\end{equation}
%\end{widetext}

\begin{equation} \label{MP7gamma-a}
	\begin{aligned}
		\mathcal{M}^{P(a)}_{7\gamma} =-\mathcal{M}^{S(a)}_{7\gamma},
	\end{aligned}
\end{equation}

\begin{equation}  \label{MP7gamma-b}
	\begin{aligned}
		\mathcal{M}^{P(b)}_{7\gamma} =-\mathcal{M}^{S(b)}_{7\gamma},
	\end{aligned}
\end{equation}
where $\eta_t=V_{tb} V_{td}^*$. The superscripts (a) and (b) mean the contributions of the diagrams (a) and (b) in Fig. \ref{fig-O7gamma}, respectively. The integration of $x_1$ is from $x^d_1$ to $x^u_1$ with $x^{u,d}_1=1/2\pm\sqrt{1/4-|\vec{k_\perp}|^2/m_B^2}$, which can keep the virtual mass of the $b$ quark in $B$ meson to be all positive. The details of this can be found in Ref. \cite{Lu-Yang2021}. The functions $H^a_{7}$ and $H^b_{7}$ are as follows:

%\begin{widetext}
\begin{equation}
	\begin{aligned}
		% --- O7 Gamma (a) ---
		& H^a_7(x_1, x_2, b_1, b_2) = K_0\left(\sqrt{x_1 x_2} m_B b_2\right) \\
		&\quad \times \left[ \theta(b_2-b_1) I_0(\sqrt{x_1} m_B b_1) K_0(\sqrt{x_1} m_B b_2) \right.\\
&+ \left.\theta(b_1-b_2) I_0(\sqrt{x_1} m_B b_2) K_0(\sqrt{x_1} m_B b_1) \right], \\
		% --- O7 Gamma (b) ---
          &H^b_7(x_1, x_2, b_1, b_2) =H^a_7(x_2, x_1, b_2, b_1).
	%	&H^b_7(x_1, x_2, b_1, b_2) = K_0\left(\sqrt{x_1 x_2} m_B b_1\right) \\
	%	&\quad \times \left[ \theta(b_2-b_1) I_0(\sqrt{x_2} m_B b_1) K_0(\sqrt{x_2} m_B b_2) \right.\\
      %&+ \left.\theta(b_1-b_2) I_0(\sqrt{x_2} m_B b_2) K_0(\sqrt{x_2} m_B b_1) \right].
	\end{aligned}
\end{equation}
%\end{widetext}
Here $K_0$, $I_0$ are the modified Bessel functions. $t^a_7$ and $t^b_7$ in Eqs. (\ref{M7gamma-a}) and (\ref{M7gamma-b}) are the largest energy scale of the virtual quark and gluon in the diagrams of \ref{fig-O7gamma} (a) and (b), which are
\begin{align}
	% --- O7 Gamma ---
	% th1r1
	t^a_{7} &= \max\left( \sqrt{x_1} m_B, \sqrt{x_1 x_2} m_B, \frac{1}{b_1}, \frac{1}{b_2} \right) \\
	% th1r2
	t^b_{7} &= \max\left( \sqrt{x_2} m_B, \sqrt{x_1 x_2} m_B, \frac{1}{b_1}, \frac{1}{b_2} \right) 
\end{align}
The functions $S_B$ and $S_V$ are functions related to the Sudakov factor, and $S_t$ the threshold factor, which can be found in Appendix \ref{Appendix A}.

\begin{figure}[htbp]
	\centering
	% 直接插入刚才生成的 PDF 文件
	\includegraphics[width=7.5cm]{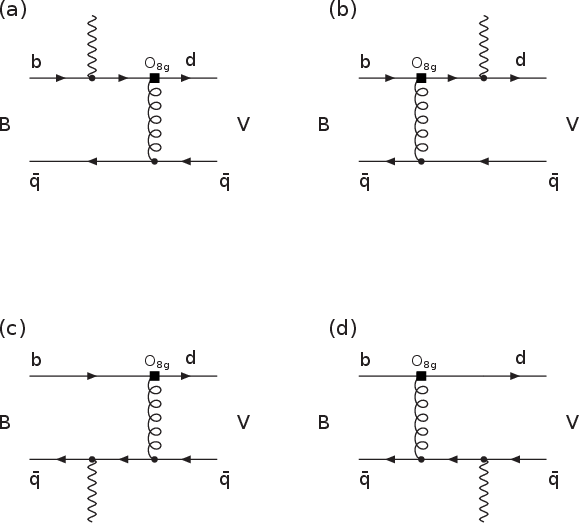}
	\caption{Contribution of operator $O_{8g}$ to
		$B\to\rho (\omega )\gamma $ decay.}
	\label{figO8g}%\label{fig:feynman1}
\end{figure}

The diagrams for the operator $O_{8g}$ are shown in Fig. \ref{figO8g}, where gluons from the operator $O_{8g}$ are connected to all the quark lines. The contributions of $O_{8g}$ to $M^S$ and $M^P$ are

% Diagram (a) - th1r3
\begin{equation}
	\begin{aligned}\label{O8gsa}
		&\mathcal{M}^{S(a)}_{8g} =  \frac{e m_B^4 f_B G_F}{\sqrt{27}} \eta_t Q_{q}\int_0^\infty k_T dk_T  \int_{x_1^d}^{x_1^u} dx_1 \\
&\times\int_0^1 dx_2\int_0^{1/\Lambda} b_1 db_1 b_2 db_2 \, \alpha_s(t^a_8) \, C_{8g}(t^a_8)\left( \frac{m_B}{2}\right. \\
		&  +\left. \frac{k_T^2}{2 x_1^2 m_B} \right) (E_Q + m_Q) \{ - x_2 (E_q - k_z) m_0 \\
&\times\left[ \phi_V^a(x_2, b_2) + \phi_V^v(x_2, b_2) \right] \} K(k) \, J_0(k_T b_1)  \\
		& \, H^a_8(x_1, x_2, b_1, b_2) \exp\left[ -S_B(t^a_8) - S_V(t^a_8) \right] \\
&\times\, S_t(x_1),
	\end{aligned}
\end{equation}

% Diagram (b) - th1r4, th1r5
\begin{equation}
	\begin{aligned}
		&\mathcal{M}^{S(b)}_{8g} =  \frac{e m_B^4 f_B G_F}{\sqrt{27}} \eta_t Q_{q}\int_0^\infty k_T dk_T  \int_{x_1^d}^{x_1^u} dx_1\\
& \times\int_0^1 dx_2 \int_0^{1/\Lambda} b_1 db_1 b_2 db_2 \, \alpha_s(t^b_8) \, C_{8g}(t^b_8) \left( \frac{m_B}{2}\right.\\
		&  + \left.\frac{k_T^2}{2 x_1^2 m_B} \right) (E_Q + m_Q) x_2 \bigg\{ m_0 (3E_q + k_z) \\
&\times [ \phi_V^a(x_2, b_2) + \phi_V^v(x_2, b_2) ]  + 2 m_B (k_z - E_q) \\
& \times \phi_V^T(x_2, b_2) \bigg\}K(k) \, J_0(k_T b_1) \, H^b_8(x_1, x_2, b_1, b_2) \\
\end{aligned}\nonumber
\end{equation}
\begin{equation}
\begin{aligned}&\times\exp\left[ -S_B(t^b_8) - S_V(t^b_8) \right] \, S_t(x_2),
	\end{aligned}
\end{equation}

% Diagram (c) - th1r6, th1r7
\begin{equation}
	\begin{aligned}
		&\mathcal{M}^{S(c)}_{8g} =  \frac{e m_B^4 f_B G_F}{\sqrt{27}} \eta_t Q_{q}\int_0^\infty k_T dk_T \int_{x_1^d}^{x_1^u} dx_1 \\
&\times\int_0^1 dx_2 \int_0^{1/\Lambda} b_1 db_1 b_2 db_2  \, \alpha_s(t^c_8) \, C_{8g}(t^c_8) \bigg( \frac{m_B}{2}  \\
		& + \frac{k_T^2}{2 x_1^2 m_B} \bigg) (E_Q + m_Q) \{ x_2 (E_q + k_z)m_0\\
&\times  \left[ \phi_V^a(x_2, b_2) + \phi_V^v(x_2, b_2) \right] \} K(k) \, J_0(k_T b_1)\\
		& \times  \, H^c_8(x_1, x_2, b_1, b_2) \exp\left[ -S_B(t^c_8) - S_V(t^c_8) \right] \\
&\times S_t(x_1),
	\end{aligned}
\end{equation}

% Diagram (d) - th1r8, th1r9
\begin{equation}
	\begin{aligned}
		&\mathcal{M}^{S(d)}_{8g} =  \frac{e m_B^4 f_B G_F}{\sqrt{27}} \eta_t Q_{q} \int_0^\infty k_T dk_T \int_{x_1^d}^{x_1^u} dx_1\\
&\times\int_0^1 dx_2  \int_0^{1/\Lambda} b_1 db_1 b_2 db_2 \, \alpha_s(t^d_8) \, C_{8g}(t^d_8) \bigg( \frac{m_B}{2}\\
		&  + \frac{k_T^2}{2 x_1^2 m_B} \bigg) (E_Q + m_Q) \bigg\{ m_0 x_2 [ 6 E_q \phi_V^v(x_2, b_2) \\
&- 2 k_z \phi_V^a(x_2, b_2) ]+ m_B \left[ E_q(x_2+2) - k_z(x_2-2) \right] \\
		& \times  \phi_V^T(x_2, b_2) \bigg\}K(k) \, J_0(k_T b_1) \, H^d_8(x_1, x_2, b_1, b_2) \\
		& \times  \exp\left[ -S_B(t^d_8) - S_V(t^d_8) \right] \, S_t(x_2),
	\end{aligned}
\end{equation}

 \begin{equation}
	\begin{aligned}
		\mathcal{M}^{P(a)}_{8g} =-\mathcal{M}^{S(a)}_{8g},
	\end{aligned}
\end{equation}

\begin{equation}
	\begin{aligned}
		\mathcal{M}^{P(b)}_{8g} =-\mathcal{M}^{S(b)}_{8g},
	\end{aligned}
\end{equation}

\begin{equation}
	\begin{aligned}
		\mathcal{M}^{P(c)}_{8g} =-\mathcal{M}^{S(c)}_{8g},
	\end{aligned}
\end{equation}

\begin{equation}
	\begin{aligned}
		&\mathcal{M}^{P(d)}_{8g} =  \frac{e m_B^4 f_B G_F}{\sqrt{27}} \eta_t Q_{q}\int_0^\infty k_T dk_T  \int_{x_1^d}^{x_1^u} dx_1\\
&\times \int_0^1 dx_2 \int_0^{1/\Lambda} b_1 db_1 b_2 db_2 \, \alpha_s(t^d_8) \, C_{8g}(t^d_8) \\
	\end{aligned}\nonumber
\end{equation}
\begin{equation}
	\begin{aligned}\label{O8gpd}
		& \times \left( \frac{m_B}{2} + \frac{k_T^2}{2 x_1^2 m_B} \right) (E_Q + m_Q) \bigg\{ m_0 x_2 \\
&\times\left[ 6 E_q \phi_V^a(x_2, b_2) - 2 k_z \phi_V^v(x_2, b_2) \right]+ m_B  \\
		& \times \left[ E_q(x_2+2) - k_z(x_2-2) \right] \phi_V^T(x_2, b_2) \bigg\} \\
		& \times K(k) \, J_0(k_T b_1) \, H^d_8(x_1, x_2, b_1, b_2)\\
&\times \exp\left[ -S_B(t^d_8) - S_V(t^d_8) \right] \, S_t(x_2),
	\end{aligned}
\end{equation}
% Helper notation for brevity (optional, put in text)
% H_0^{(1)} = J_0 + i Y_0 is the Hankel function of the first kind.
% (b_1 <-> b_2) implies the symmetric term for b_1 > b_2.

\begin{equation}
	\begin{aligned}
		% --- O8g (a) H^a_{8g}---
		&H^a_8(x_1, x_2, b_1, b_2) = K_0\left(\sqrt{x_1 x_2} m_B b_2\right) \left[ \theta(b_2-b_1) \right.\\
		&\quad \times I_0(\sqrt{x_1+1} m_B b_1) K_0(\sqrt{x_1+1} m_B b_2) \\
&\quad+ \theta(b_1-b_2) I_0(\sqrt{x_1+1} m_B b_2) \\
&\quad\times K_0(\sqrt{x_1+1} m_B b_1) \left.\right] \\
	\end{aligned}
\end{equation}
		% --- O8g (b)H^b_{8g} ---
\begin{equation}
	\begin{aligned}
		&H^b_8(x_1, x_2, b_1, b_2) = i \frac{\pi}{2} K_0\left(\sqrt{x_1 x_2} m_B b_1\right)[ \theta(b_2-b_1) \\
&\quad\times J_0(\sqrt{1-x_2} m_B b_1) H_0^{(1)}(\sqrt{1-x_2} m_B b_2)\\
		&\quad    + \theta(b_1-b_2) J_0(\sqrt{1-x_2} m_B b_2) \\
&\quad\times H_0^{(1)}(\sqrt{1-x_2} m_B b_1) ],
	\end{aligned}
\end{equation}

% --- O8g (c) Unified H^c_{8g}---

\begin{equation}
	\begin{aligned}
		& H^c_8(x_1, x_2, b_1, b_2)= \left[\right. \theta(x_1-x_2) K_0\left(\sqrt{x_1-x_2} m_B b_2\right) \\
&\quad + \theta(x_2-x_1) \frac{i\pi}{2} H_0^{(1)}\left(\sqrt{x_2-x_1} m_B b_2\right)\left. \right] \\
		& \quad \times \left[ \right.\theta(b_2-b_1) I_0(\sqrt{x_1} m_B b_1) K_0(\sqrt{x_1} m_B b_2) \\
&\quad+ \theta(b_1-b_2) I_0(\sqrt{x_1} m_B b_2) K_0(\sqrt{x_1} m_B b_1)\left. \right],
	\end{aligned}
\end{equation}

\begin{equation}
	\begin{split}
		H^d_8&(x_1, x_2, b_1, b_2) = \\
		&\left[ \theta(x_1-x_2) \frac{i\pi}{2} K_0\left(\sqrt{x_1-x_2} m_B b_1\right) \right. \\
		&\left. - \theta(x_2-x_1) \frac{\pi^2}{4} H_0^{(1)}\left(\sqrt{x_2-x_1} m_B b_1\right) \right] \\
		&\times \left[ \theta(b_2-b_1) J_0(\sqrt{x_2} m_B b_1) H_0^{(1)}(\sqrt{x_2} m_B b_2) \right. \\
		&\left. + \theta(b_1-b_2) J_0(\sqrt{x_2} m_B b_2) H_0^{(1)}(\sqrt{x_2} m_B b_1) \right],
	\end{split}
\end{equation}
where $H_0^{(1)}$ is the Hankel function of the first kind, and

\begin{align}
	% --- O8g ---
	% th1r3
	t^a_{8g} &= \max\left( \sqrt{1+x_1} m_B, \sqrt{x_1 x_2} m_B, \frac{1}{b_1}, \frac{1}{b_2} \right), \\
	% th1r4, th1r5
	t^b_{8g} &= \max\left( \sqrt{1-x_2} m_B, \sqrt{x_1 x_2} m_B, \frac{1}{b_1}, \frac{1}{b_2} \right), \\
	% th1r6, th1r7 (Note: Combined x1-x2 and x2-x1 logic into absolute value)
	t^c_{8g} &= \max\left( \sqrt{x_1} m_B, \sqrt{|x_1-x_2|} m_B, \frac{1}{b_1}, \frac{1}{b_2} \right), \\
	% th1r8, th1r9, th1r10, th1r11
	t^d_{8g} &= \max\left( \sqrt{x_2} m_B, \sqrt{|x_1-x_2|} m_B, \frac{1}{b_1}, \frac{1}{b_2} \right),
\end{align}

Using the results of the amplitudes of $M^{j(a)}_{7\gamma}$ and $M^{j(b)}_{7\gamma}$ ($j=S, P$) given by Eqs. (\ref{M7gamma-a}) $\sim$ (\ref{MP7gamma-b}), which are contributed by the diagrams (a) and (b) in Fig. \ref{fig-O7gamma}, the amplitudes for each decay modes caused by the operator $O_{7\gamma}$ are 
\begin{align}
	M(B^+ \to \rho^+ \gamma)_{7\gamma}^j &= M_{7\gamma}^{j(a)} + M_{7\gamma}^{j(b)}, \\
	M(B^0 \to \rho^0 \gamma)_{7\gamma}^j &= -\frac{1}{\sqrt{2}} \left[ M_{7\gamma}^{j(a)} + M_{7\gamma}^{j(b)} \right], \\
	M(B^0 \to \omega \gamma)_{7\gamma}^j &= \frac{1}{\sqrt{2}} \left[ M_{7\gamma}^{j(a)} + M_{7\gamma}^{j(b)} \right].\end{align}
The amplitudes caused by the operator $O_{8g}$ which are corresponding to Fig. \ref{figO8g} are in the following
\begin{align}
	\begin{split}
		&M(B^+ \to \rho^+ \gamma)_{8g}^j = M_{8g}^{j(a)}(Q_b)  \\
		&\quad + M_{8g}^{j(b)}(Q_d)+ M_{8g}^{j(c)}(Q_u) + M_{8g}^{j(d)}(Q_u),
	\end{split} \\
	\begin{split}
		 &M(B^0 \to \rho^0 \gamma)_{8g}^j= -\frac{1}{\sqrt{2}} \Bigl[ M_{8g}^{j(a)}(Q_b)  \\
		&\quad + M_{8g}^{j(b)}(Q_d)+ M_{8g}^{j(c)}(Q_d) + M_{8g}^{j(d)}(Q_d) \Bigr],
	\end{split} \\
	\begin{split}
		&M(B^0 \to \omega \gamma)_{8g}^j = \frac{1}{\sqrt{2}} \Bigl[ M_{8g}^{j(a)}(Q_b)  \\
		&\quad + M_{8g}^{j(b)}(Q_d)+ M_{8g}^{j(c)}(Q_d) + M_{8g}^{j(d)}(Q_d) \Bigr],
	\end{split}
\end{align}
where $M^{j(a,b,c,d)}_{8g}(Q_q)$, $(j=S,P)$, can be found in Eqs. (\ref{O8gsa}) $\sim$ (\ref{O8gpd}).
	% 图片示例 (请确保目录下有 figure.pdf 文件，否则注释掉)
	% \begin{figure}[htbp]
		%     \centering
		%     \includegraphics[width=0.9\linewidth]{figure.pdf}
		%     \caption{This is the caption for the figure.}
		%     \label{fig:example}
		% \end{figure}
	
\section{ THE CONTRIBUTION OF NEXT-TO-LEADING-ORDER CORRECTIONS} \label{sec-NLO}
\begin{figure}[htbp]
 	\centering
 	% 直接插入刚才生成的 PDF 文件
 	\includegraphics[width=7.5cm]{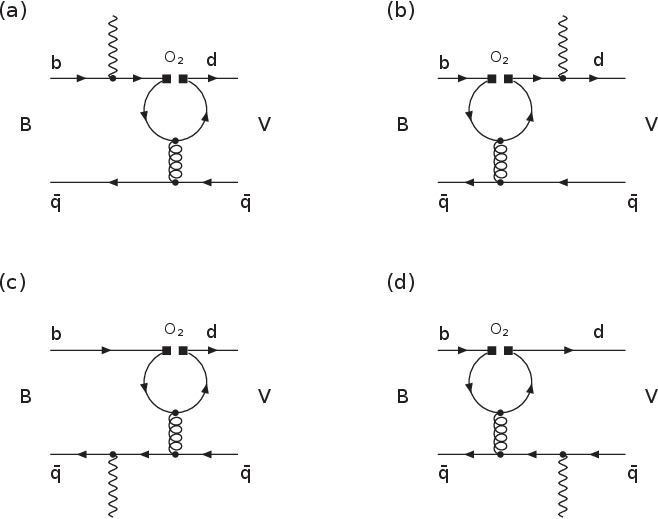}
 \caption{Diagrams in which the operator $O_2$ is inserted in the loop,and a photon is emitted from the external quark line.}\label{fig-loop-a}
 	%\label{fig:feynman1}
 \end{figure}

 In this section, we consider the contributions of the diagrams with the effective operators $O_{i}$'s being inserted in the quark loops. These diagrams can be divided into two types. One type is for the diagrams where the photons are emitted from the external quark lines, which are shown in Fig. \ref{fig-loop-a}. The other type is for the diagrams with the photons being emitted from the internal quark lines in the loop, which are shown in Fig. \ref{fig-loop-b}. In this work we only consider the insertion of the tree-level operators $O_1$ and $O_2$, and neglect the penguin operators $O_{3\sim 6}$, because their contributions are further suppressed by their small Wilson coefficients in addition to the loop suppression. For the tree-level operators $O_1$ and $O_2$, only $O_2$ contributes because the color of $O_1$ does not match the quark loops with one gluon attached on it.
 
\subsection{Contributions of external-quark-line emission}

The function for the inner quark loop is 
\begin{equation} \label{I-nu}
\begin{split}
	I^\nu &= \frac{g}{8\pi^2} \left[ \frac{2}{3} - G(m_i^2, k^2, \mu) \right] \bar{b} T^a_{ij}(k^2 \gamma^\nu \\
&\quad  - k^\nu \not{k}) (1 - \gamma_5) d
\end{split}
\end{equation}
with $G(m_i^2, k^2, \mu)$ is  
\begin{equation}
	\begin{split}
		&G(m_i^2, k^2, \mu) = - \int_{0}^{1} dx 4x(1-x) \\
		&\quad \times \log\left[\frac{m_i^2 - x(1-x)k^2 - i\epsilon}{\mu^2}\right],
	\end{split}
\end{equation}
where $m_i$ is the mass of the quark in the inner quark loop, and here $i=u$ and $c$ quarks. With the function $I^\nu$ given in Eq. (\ref{I-nu}), the contributions of the diagrams (a)$\sim$ (d) in Fig. \ref{fig-loop-a} are calculated to be
% Diagram (a)
\begin{widetext}
% Diagram (a)
\begin{equation}
	\begin{aligned}
		\mathcal{M}^{(a)}_{1i} =& \frac{1}{2} \frac{e m_B^4 f_B G_F}{\sqrt{27}} \eta_i Q_{q}\int_0^\infty k_T dk_T  \int_{x_1^d}^{x_1^u} dx_1\int_0^1 dx_2  \int_0^{1/\Lambda} b_1 db_1 b_2 db_2 \, \alpha_s(t^a_8) \, C_2(t^a_8) \\
		& \times \left( \frac{m_B}{2} + \frac{k_T^2}{2 x_1^2 m_B} \right) (E_q + k_z)(E_Q + m_Q) \, x_1 x_2 \, m_0 \left[ \phi_V^a(x_2, b_2) - \phi_V^v(x_2, b_2) \right] \\
		& \times \left[  G\left(m_i^2, -\frac{m_B^2}{4}, t^a_8\right) - \frac{2}{3} \right] K(k) \, J_0(k_T b_1) \, H^a_8(x_1, x_2, b_1, b_2)\exp\left[ \right.-S_B(t^a_8)\\
		&   - S_V(t^a_8) \left.\right] \,S_t(x_1),
	\end{aligned}
\end{equation}

% Diagram (b)
\begin{equation}
	\begin{aligned}
		\mathcal{M}^{(b)}_{1i} =& \frac{1}{2} \frac{e m_B^4 f_B G_F}{\sqrt{27}} \eta_i Q_{q}\int_0^\infty k_T dk_T  \int_{x_1^d}^{x_1^u} dx_1\int_0^1 dx_2  \int_0^{1/\Lambda} b_1 db_1 b_2 db_2 \, \alpha_s(t^b_8) \, C_2(t^b_8) \\
		& \times \left( \frac{m_B}{2} + \frac{k_T^2}{2 x_1^2 m_B} \right) (E_q - k_z)(E_Q + m_Q) \, x_2^2 \, m_0 \left[ \phi_V^a(x_2, b_2) + \phi_V^v(x_2, b_2) \right] \\
		& \times \left[  G\left(m_i^2, -\frac{m_B^2}{4}, t^b_8\right) - \frac{2}{3} \right]  K(k) \, J_0(k_T b_1) \, H^b_8(x_1, x_2, b_1, b_2)\exp\left[ \right.-S_B(t^b_8)\\
		&   - S_V(t^b_8) \left.\right] \, S_t(x_2),
	\end{aligned}
\end{equation}

% Diagram (c)
\begin{equation}
	\begin{aligned}
		\mathcal{M}^{(c)}_{1i} =& \frac{1}{2} \frac{e m_B^4 f_B G_F}{\sqrt{27}} \eta_i Q_{q}\int_0^\infty k_T dk_T  \int_{x_1^d}^{x_1^u} dx_1\int_0^1 dx_2  \int_0^{1/\Lambda} b_1 db_1 b_2 db_2 \, \alpha_s(t^c_8) \, C_2(t^c_8) \\
		& \times \left( \frac{m_B}{2} + \frac{k_T^2}{2 x_1^2 m_B} \right) (E_q + k_z)(E_Q + m_Q) \, x_2 \, m_0 \left[ \phi_V^a(x_2, b_2) + \phi_V^v(x_2, b_2) \right] \\
		& \times \left[  G\left(m_i^2, -\frac{m_B^2}{4}, t^c_8\right) - \frac{2}{3} \right]K(k) \, J_0(k_T b_1) \, H^c_8(x_1, x_2, b_1, b_2) \exp\left[\right. -S_B(t^c_8) \\
		& - S_V(t^c_8) \left.\right] \, S_t(x_1),
	\end{aligned}
\end{equation}
% Diagram (d) S-wave
\begin{equation}
	\begin{aligned}
		\mathcal{M}^{S(d)}_{1i} =& \frac{1}{2} \frac{e m_B^4 f_B G_F}{\sqrt{27}} \eta_i Q_{q} \int_0^\infty k_T dk_T\int_{x_1^d}^{x_1^u} dx_1\int_0^1 dx_2  \int_0^{1/\Lambda} b_1 db_1 b_2 db_2  \, \alpha_s(t^d_8) \, C_2(t^d_8) \\
		& \times \left( \frac{m_B}{2} + \frac{k_T^2}{2 x_1^2 m_B} \right) (E_Q + m_Q) \, x_2 \left[  G\left(m_i^2, -\frac{m_B^2}{4}, t^d_8\right) - \frac{2}{3} \right] \bigg\{ m_0 \Big[ \phi_V^a(x_2, b_2)\\
		& \times \left( E_q(x_2-1) - k_z(x_2+1) \right)+ 3 \phi_V^v(x_2, b_2)\left(\right. E_q(x_2+1) - k_z(x_2-1) \left.\right) \Big] \\
		& \quad + m_B \phi_V^T(x_2, b_2) (3 E_q + k_z) \bigg\} K(k) \, J_0(k_T b_1) \, H^d_8(x_1, x_2, b_1, b_2)\exp\left[\right. -S_B(t^d_8)  \, \\
		& - S_V(t^d_8) \left.\right]  S_t(x_2),
	\end{aligned}
\end{equation}
\begin{equation}
	\begin{aligned}
		\mathcal{M}^{P(a)}_{1i} =\mathcal{M}^{S(a)}_{1i},\quad \mathcal{M}^{P(b)}_{1i} =-\mathcal{M}^{S(b)}_{1i},
            \quad \mathcal{M}^{P(c)}_{1i} =-\mathcal{M}^{S(c)}_{1i},
	\end{aligned}
\end{equation}

% Diagram (d) P-wave
\begin{equation}
	\begin{aligned}
		\mathcal{M}^{P(d)}_{1i} =& -\frac{1}{2} \frac{e m_B^4 f_B G_F}{\sqrt{27}} \eta_i Q_{q} \int_0^\infty k_T dk_T\int_{x_1^d}^{x_1^u} dx_1\int_0^1 dx_2  \int_0^{1/\Lambda} b_1 db_1 b_2 db_2  \, \alpha_s(t^d_8) \, C_2(t^d_8) \\
		& \times \left( \frac{m_B}{2} + \frac{k_T^2}{2 x_1^2 m_B} \right) (E_Q + m_Q) \, x_2 \left[  G\left(m_i^2, -\frac{m_B^2}{4}, t^d_8\right) - \frac{2}{3} \right] \bigg\{ m_0 \Big[ 3 \phi_V^a(x_2, b_2)\\
		& \times \left( E_q(x_2+1) - k_z(x_2-1) \right)  + \phi_V^v(x_2, b_2)\left( E_q(x_2-1) - k_z(x_2+1) \right) \Big] \\
		& \quad + m_B \phi_V^T(x_2, b_2) (3 E_q + k_z) \bigg\}  K(k) \, J_0(k_T b_1) \, H^d_8(x_1, x_2, b_1, b_2) \exp\left[ \right.-S_B(t^d_8)\\
		&  - S_V(t^d_8) \left.\right] \, S_t(x_2),
	\end{aligned}
\end{equation}
where the subscript $i$ indicates the flavor of the inner quarks propagating in the quark loop.

\end{widetext}
The amplitudes of $B^+ \to \rho^+ \gamma$, $B^0 \to \rho^0 \gamma$ and $B^0 \to \omega \gamma$ contributed by the inner quark-loop diagrams are
\begin{align}
	M(B^+ \to \rho^+ \gamma)_{1i}^j &= M_{1i}^{j(a)}(Q_b) + M_{1i}^{j(b)}(Q_d)  \notag \\
	&\quad + M_{1i}^{j(c)}(Q_u)+ M_{1i}^{j(d)}(Q_u),  \\[10pt]
	M(B^0 \to \rho^0 \gamma)_{1i}^j &= -\frac{1}{\sqrt{2}} \Big[ M_{1i}^{j(a)}(Q_b) + M_{1i}^{j(b)}(Q_d) \notag \\
	&\quad + M_{1i}^{j(c)}(Q_d) + M_{1i}^{j(d)}(Q_d) \Big],  \\[10pt]
	M(B^0 \to \omega \gamma)_{1i}^j &= \frac{1}{\sqrt{2}} \Big[ M_{1i}^{j(a)}(Q_b) + M_{1i}^{j(b)}(Q_d) \notag \\
	&\quad  + M_{1i}^{j(c)}(Q_d)+ M_{1i}^{j(d)}(Q_d) \Big]. 
\end{align}

\subsection{Contributions of internal-loop-quark-line emission}

 \begin{figure}[htbp]
	\centering
	% 直接插入刚才生成的 PDF 文件
	\includegraphics[width=8.0cm]{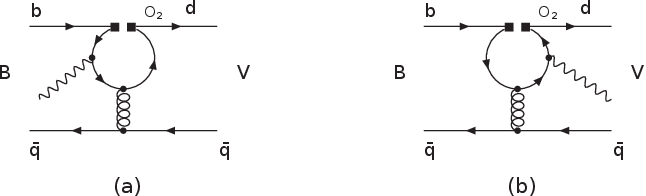}
	\caption{Diagrams in which the operator $O_2$ is inserted in the quark 
		loop,and a photon is emitted from the internal loop quark line.}
	\label{fig-loop-b}
      %\label{fig:feynman1}
\end{figure}
The effective interaction term for $b\to d$ transition with one photon and one gluon emitted from the quark loop can be calculated in the $\overline{MS}$ scheme, which is
\begin{equation}
	I = \bar{d} \gamma^\rho (1 - \gamma_5) T^a b I_{\mu\nu\rho} \varepsilon^\mu_\gamma \varepsilon^\nu_g,\label{57}
\end{equation}
with
\begin{align}
	I_{\mu\nu\rho} &= A_4 \Big[ (q \cdot k) \epsilon_{\mu\nu\rho\sigma} (q-k)^\sigma + \epsilon_{\nu\rho\sigma\tau} q^\sigma k^\tau k_\mu \notag \\
	&\quad - \epsilon_{\mu\rho\sigma\tau} q^\sigma k^\tau q_\nu \Big] + A_5 \Big[ \epsilon_{\mu\rho\sigma\tau} q^\sigma k^\tau k_\nu \notag \\
	&\quad - k^2 \epsilon_{\mu\nu\rho\sigma} q^\sigma \Big], \tag{58} %\\[10pt]
\end{align}
where $q$ is the momentum of the photon, $k$ the momentum of the gluon, and

\begin{align}
	A_4 &= \frac{4ieg}{3\pi^2} \int_0^1 dx \int_0^{1-x} dy \notag \\
	&\quad \times \frac{xy}{x(1-x)k^2 + 2xyqk - m_i^2 + i\varepsilon}, \tag{59} \\[10pt]
	A_5 &= -\frac{4ieg}{3\pi^2} \int_0^1 dx \int_0^{1-x} dy \notag \\
	&\quad \times \frac{x(1-x)}{x(1-x)k^2 + 2xyqk - m_i^2 + i\varepsilon}. \tag{60}
\end{align}

With the effective interaction given in Eq. (\ref{57}), the amplitudes contributed by the diagrams (a) and (b) shown in  Fig. \ref{fig-loop-b} can be calculated, and the results are
\begin{widetext}
% Loop E (S-wave)
\begin{equation}
	\begin{aligned}
		\mathcal{M}^{S}_{2i} =& \frac{16}{3} \frac{e m_B^4 f_B G_F}{\sqrt{27}} \eta_i\int_0^\infty k_T dk_T  \int_{x_1^d}^{x_1^u} dx_1\int_0^1 dx_2  \int_0^1 dy_1 dy_2 \int_0^{1/\Lambda} b_1 db_1 \, \alpha_s(t_{2i}) \, C_2(t_{2i}) \\
		& \times \left( \frac{m_B}{2} + \frac{k_T^2}{2 x_1^2 m_B} \right) (E_Q + m_Q) \frac{x_2}{y_1 y_2 x_2 m_B^2 - m_i^2}  \bigg\{ - y_1(1-y_1) x_2 (E_q - k_z) m_0 \\
		& \times\left[ \phi_V^a(x_2, b_1) + \phi_V^v(x_2, b_1) \right]- y_1 y_2 \Big[ m_0 \left( \right.(E_q+k_z)\phi_V^a(x_2, b_1)+ (k_z(1+2x_2) \\
		& \quad   + E_q(1-2x_2))\phi_V^v(x_2, b_1)\left. \right)  + m_B (k_z - E_q) \phi_V^T(x_2, b_1) \Big] \bigg\} K(k) \, J_0(k_T b_1) \,\\
		& \times  H^{(e)}_{2i}(x_1, x_2, y_1, y_2, b_1) \exp\left[ -S_B(t_{2i}) - S_V(t_{2i}) \right],
	\end{aligned}
\end{equation}

% Loop E (P-wave)
\begin{equation}
	\begin{aligned}
		\mathcal{M}^{P}_{2i} =& -\frac{16}{3} \frac{e m_B^4 f_B G_F}{\sqrt{27}} \eta_i \int_0^\infty k_T dk_T \int_{x_1^d}^{x_1^u} dx_1\int_0^1 dx_2  \int_0^1 dy_1 dy_2 \int_0^{1/\Lambda} b_1 db_1 \, \alpha_s(t_{2i}) \, C_2(t_{2i}) \\
		& \times \left( \frac{m_B}{2} + \frac{k_T^2}{2 x_1^2 m_B} \right) (E_Q + m_Q) \frac{x_2}{y_1 y_2 x_2 m_B^2 - m_i^2}\bigg\{ y_1(1-y_1) x_2 (E_q - k_z) m_0 \\
		& \times  \left[ \phi_V^a(x_2, b_1) + \phi_V^v(x_2, b_1) \right] + y_1 y_2 \Big[ m_0 \left( \right.(E_q+k_z)\phi_V^v(x_2, b_1) + (k_z(1+2x_2) \\
		& \quad + E_q(1-2x_2))\phi_V^a(x_2, b_1) \left.\right) + m_B (k_z - E_q) \phi_V^T(x_2, b_1) \Big] \bigg\} K(k) \, J_0(k_T b_1) \,\\
		& \times  H^{(e)}_{2i}(x_1, x_2, y_1, y_2, b_1) \exp\left[ -S_B(t_{2i}) - S_V(t_{2i}) \right],
	\end{aligned}
\end{equation}
\end{widetext}
where
\begin{align}
	% Definition of x_yc
	x_{yi} &= x_1 x_2 m_B^2 - \frac{y_2 x_2 m_B^2}{1-y_1} + \frac{m_i^2}{y_1(1-y_1)},% \\[10pt]
\end{align}
	% Unified Hard Kernel H_2i
and the hard function and hard scale are 
\begin{align}
	H^{(e)}_{2i} &= \left[ \right. \theta(x_{yi}) K_0\left(\sqrt{x_{yi}} b_1\right) + \theta(-x_{yi}) \frac{i\pi}{2}\nonumber\\
 &\times H_0^{(1)}\left(\sqrt{-x_{yi}} b_1\right) \left.\right] - K_0\left(\sqrt{x_1 x_2} m_B b_1\right), \\[10pt]
	% Hard Scale t_2i
	t_{2i} &= \max\left( \sqrt{x_1 x_2} m_B, \sqrt{|x_{yi}|}, \frac{1}{b_1} \right),
\end{align}
respectively. The decay amplitudes for each decay modes are

\begin{equation}
	\begin{gathered}
		M(B^+ \to \rho^+ \gamma)_{2i}^j = M_{2i}^j,\\ %\qquad 
		M(B^0 \to \rho^0 \gamma)_{2i}^j = -\frac{M_{2i}^j}{\sqrt{2}}, \\
		M(B^0 \to \omega \gamma)_{2i}^j = \frac{M_{2i}^j}{\sqrt{2}}.
	\end{gathered}
\end{equation}

\section{The annihilation diagram contributions}\label{sec-annihilation}
In this section we consider the contributions of the annihilation diagrams, where the operators of tree-level and QCD penguins are inserted with all kinds of possibilities. 

\subsection{Tree annihilation}

\begin{figure}[htbp]
	\centering
	% 直接插入刚才生成的 PDF 文件
	\includegraphics[width=7.5cm]{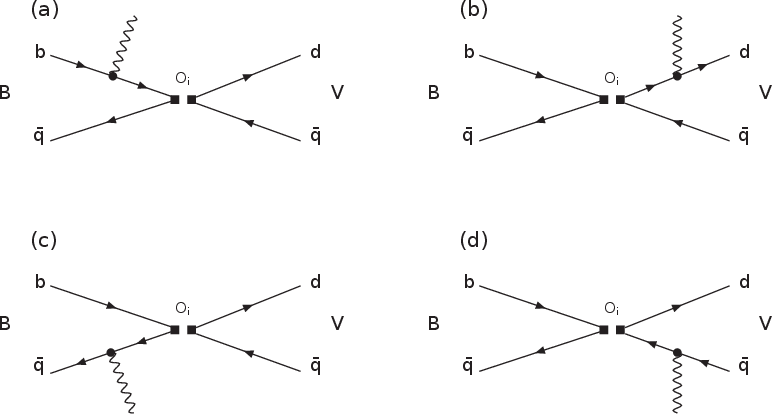}
	\caption{ Annihilation diagrams in which the operators $O_1,O_2$
		are inserted.}
	\label{fig-annihilation-a}
\end{figure}

The annihilation diagrams with the insertion of the tree-level operators are shown in Fig. \ref{fig-annihilation-a}. For the convenience of expressing the decay amplitude relevant to each diagram, we define the combinations of the tree-level Wilson coefficients as
\begin{equation}
a_1(t) = C_1(t) + \frac{C_2(t)}{3}, \qquad a_2(t) = C_2(t) + \frac{C_1(t)}{3},
\end{equation}	

then the decay amplitude for each diagram can be given as 

	% 请确保删掉了前面的 \begin{widetext}
		
		\begin{align}
		% 1. Annihilation A
		\mathcal{M}^{S(a)}_{A_{k}} &= \int_0^\infty k_T dk_T \int_{x_1^d}^{x_1^u} dx_1 \int_0^{1/\Lambda} b_1 db_1 \notag \\
		&\quad \times \eta_u \cdot \mathcal{C}a_k(t_A^a) Q_b \exp[-S_B(t_A^a)] S_t(x_1) \notag \\
		&\quad \times \left[ 4 m_B m_0 (E_q + k_z)(E_Q + m_Q) \right] \notag \\
		&\quad \times \left( \frac{m_B}{2} + \frac{k_T^2}{2 x_1^2 m_B} \right) \notag \\
		&\quad \times K(k) J_0(k_T b_1) K_0(\sqrt{1+x_1} b_1), \tag{65} \\[10pt]
			\mathcal{M}^{P(a)}_{A_{k}} &= \mathcal{M}^{S(a)}_{A_{k}}, \tag{66} \\[10pt]
			%
			% 2. Annihilation B-Scalar
			\mathcal{M}^{S(b)}_{A_{k}} &= \int_0^1 dx_2 \int_0^{1/\Lambda} b_2 db_2 \, \eta_u \cdot \mathcal{C} a_k(t_A^b) Q_q \notag \\
			&\quad \times \exp[-S_V(t_A^b)] S_t(x_2) \phi_M(x_2) m_0 \notag \\
			&\quad \times \left[ x_2 \phi_V^a(x_2) - (x_2-2)\phi_V^v(x_2) \right] \notag \\
			&\quad \times \left( -i \frac{\pi}{2} H_0^{(1)}(\sqrt{1-x_2} b_2) \right), \tag{67} \\[10pt]
			%
			% 3. Annihilation B-Pseudoscalar
			\mathcal{M}^{P(b)}_{A_{k}} &= \int_0^1 dx_2 \int_0^{1/\Lambda} b_2 db_2 \, \eta_u \cdot \mathcal{C} a_k(t_A^b) Q_q \notag \\
			&\quad \times \exp[-S_V(t_A^b)] S_t(x_2) \phi_M(x_2) m_0 \notag \\
			&\quad \times \left[ (x_2-2) \phi_V^a(x_2) - x_2 \phi_V^v(x_2) \right] \notag \\
			&\quad \times \left( -i \frac{\pi}{2} H_0^{(1)}(\sqrt{1-x_2} b_2) \right), \tag{68} \\[10pt]
			%
		% 4. Annihilation C
		\mathcal{M}^{S(c)}_{A_{k}} &= \int_0^\infty k_T dk_T\int_{x_1^d}^{x_1^u} dx_1 \int_0^{1/\Lambda} b_1 db_1 \notag \\
		&\quad \times \eta_u \cdot \mathcal{C}a_k(t_A^c) Q_{q'} \exp[-S_B(t_A^c)] S_t(x_1) \notag \\
		&\quad \times \left[ 4 m_B m_0 (E_q + k_z)(E_Q + m_Q) \right] \notag \\
		&\quad \times \left( \frac{m_B}{2} + \frac{k_T^2}{2 x_1^2 m_B} \right) \notag \\
		&\quad \times K(k) J_0(k_T b_1) K_0(\sqrt{x_1} b_1), \tag{69} \\[10pt]
			\mathcal{M}^{P(c)}_{A_{k}} &= -\mathcal{M}^{S(c)}_{A_{k}}, \tag{70} \\[10pt]
			%
			% 5. Annihilation D-Scalar
			\mathcal{M}^{S(d)}_{A_{k}} &= \int_0^1 dx_2 \int_0^{1/\Lambda} b_2 db_2 \, \eta_u \cdot \mathcal{C} a_k(t_A^d) Q_{q''} \notag \\
			&\quad \times \exp[-S_V(t_A^d)] S_t(x_2) \phi_M(x_2) m_0 \notag \\
			&\quad \times \left[ (x_2-1)\phi_V^a(x_2) + (x_2+1)\phi_V^v(x_2) \right] \notag \\
			&\quad \times \left( -i \frac{\pi}{2} H_0^{(1)}(\sqrt{x_2} b_2) \right), \tag{71} \\[10pt]
			%
			% 6. Annihilation D-Pseudoscalar
			\mathcal{M}^{P(d)}_{A_{k}} &= \int_0^1 dx_2 \int_0^{1/\Lambda} b_2 db_2 \, \eta_u \cdot \mathcal{C} a_k(t_A^d) Q_{q''} \notag \\
			&\quad \times \exp[-S_V(t_A^d)] S_t(x_2) \phi_M(x_2) (-m_0) \notag \\
			&\quad \times \left[ (x_2+1)\phi_V^a(x_2) + (x_2-1)\phi_V^v(x_2) \right] \notag \\
			&\quad \times \left( -i \frac{\pi}{2} H_0^{(1)}(\sqrt{x_2} b_2) \right) \tag{72}
		\end{align}
		
		% 请确保删掉了后面的 \end{widetext}
where $k=1\;,2$ and $\mathcal{C}$ is
		\begin{align*}
			\mathcal{C} &= \frac{\pi f_V}{8 m_B \sqrt{6}} \frac{e m_B^2 f_B G_F}{\sqrt{27}}.
		\end{align*}	
The amplitudes for all the decay modes contributed by the annihilation diagrams with tree-level operator insertion are
	
	% 第一部分：衰变幅值公式
	% 使用 split 环境将折行的公式包裹起来，使它们共用一个编号
	\begin{align}
		\begin{split}
			&M(B^+ \to \rho^+\gamma)^j_A = M^{j(a)}_{A_2}(Q_b) + M^{j(b)}_{A_2}(Q_d) \\
			&\quad\quad + M^{j(c)}_{A_2}(Q_u) + M^{j(d)}_{A_2}(Q_u),
		\end{split} \\[10pt] % [10pt] 用于增加段落间距，防止拥挤
		\begin{split}
			&M(B^0 \to \rho^0\gamma)^j_A = \frac{1}{\sqrt{2}} \Bigl[ M^{j(a)}_{A_1}(Q_b) + M^{j(b)}_{A_1}(Q_u) \\
			&\quad\quad + M^{j(c)}_{A_1}(Q_d) + M^{j(d)}_{A_1}(Q_u) \Bigr],
		\end{split} \\[10pt]
		\begin{split}
			&M(B^0 \to \omega\gamma)^j_A = \frac{1}{\sqrt{2}} \Bigl[ M^{j(a)}_{A_1}(Q_b) + M^{j(b)}_{A_1}(Q_u) \\
			&\quad\quad + M^{j(c)}_{A_1}(Q_d) + M^{j(d)}_{A_1}(Q_u) \Bigr].
		\end{split}
	\end{align}
\begin{figure}[htbp]
	\centering
	% 直接插入刚才生成的 PDF 文件
	\includegraphics[width=7.5cm]{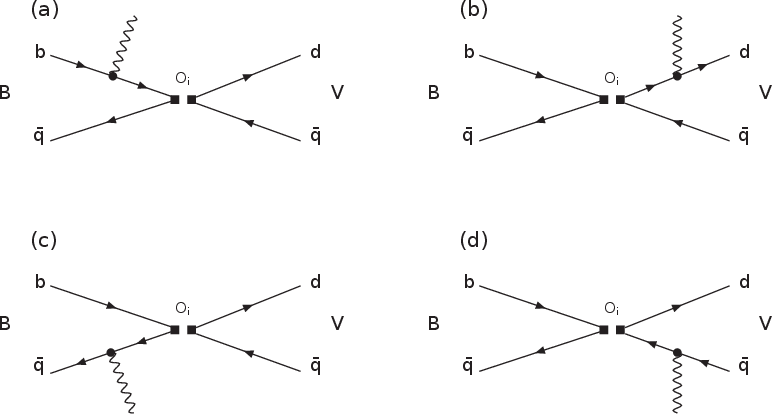}
	\caption{Annihilation diagrams in which the QCD penguin
		operators are inserted.}
	\label{fig-annihilation-b}
\end{figure}
\subsection{QCD penguin annihilation}

\begin{figure}[htbp]
	\centering
	% 直接插入刚才生成的 PDF 文件
	\includegraphics[width=7.5cm]{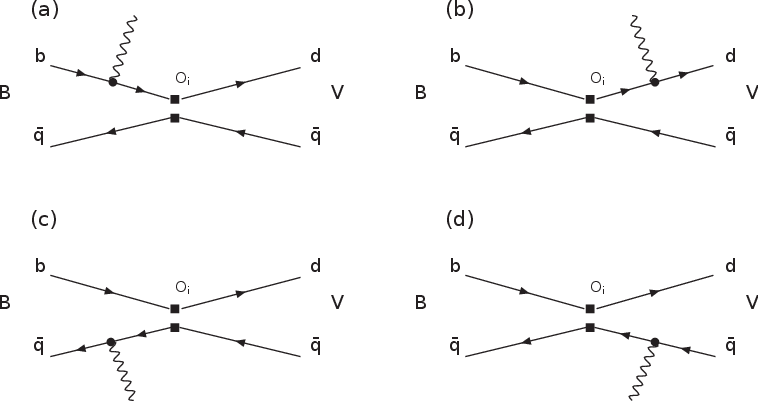}
	\caption{The other type of annihilation diagrams with operator
		insertion.}
	\label{fig-annihilation-c}
\end{figure}

The diagrams with the insertion of the penguin operators are depicted in  Fig. \ref{fig-annihilation-b} and  \ref{fig-annihilation-c}.
	% 第二部分：Wilson 系数相关公式
% 使用 & 符号来控制对齐位置，让两列公式整齐排列
% 使用 equation 开启单公式模式（只产生一个编号）

We define the combinations of the Wilson coefficients of QCD penguin operatos as
	
\begin{equation}
	% 使用 aligned 在内部进行多行、多列排版
	\begin{aligned}
		a_3(t) &= C_3(t) + \frac{C_4(t)}{3}, & \quad a_4(t) &= C_4(t) + \frac{C_3(t)}{3}, \\
		a_5(t) &= C_5(t) + \frac{C_6(t)}{3}, & \quad a_6(t) &= C_6(t) + \frac{C_5(t)}{3}.
	\end{aligned}
\end{equation}
The amplitudes for the insertion of $(V-A)(V-A)$ type of operators in Fig. \ref{fig-annihilation-b} are

	\begin{widetext}
		% 公共常数定义
%		\begin{align*}
%			\mathcal{C} &= \frac{\pi f_V}{8 m_B \sqrt{6}} \frac{e m_B^2 f_B G_F}{\sqrt{27}}
%		\end{align*}
		
		% 1. Annihilation A1 (th1r80)
		\begin{align}
			\mathcal{M}^{S(a)-}_{A1_{k}} &= \int_{x_1^d}^{x_1^u} dx_1 \int_0^{1/\Lambda} b_1 db_1 \int_0^\infty k_T dk_T \, \eta_t \cdot \mathcal{C} Q_{b} a_k(t_{A1}^a) \exp[-S_B(t_{A1}^a)] S_t(x_1) \left[\right. 4 m_B m_0 \nonumber \\
	&\quad \times (E_q + k_z)(E_Q + m_Q) \left.\right] \left( \frac{m_B}{2} + \frac{k_T^2}{2 x_1^2 m_B} \right)
			K(k) J_0(k_T b_1) K_0(\sqrt{1+x_1} b_1),
		\end{align}
		
		\begin{align}
			\mathcal{M}^{P(a)-}_{A1_{k}} &= \mathcal{M}^{S(a)-}_{A1_{k}},
		\end{align}
		
		% 2. Annihilation A1-B-Scalar (th1r81 + th1r82)
		\begin{align}
	\mathcal{M}^{S(b)-}_{A1_{k}} &= \int_0^1 dx_2 \int_0^{1/\Lambda} b_2 db_2 \, \eta_t \cdot \mathcal{C} Q_{q} 
			a_k(t_{A1}^b) \exp[-S_V(t_{A1}^b)] S_t(x_2)  \nonumber \\
			&\quad \times m_0 \left[ x_2 \phi_V^a(x_2) - (x_2-2)\phi_V^v(x_2) \right] 
			\left( -i \frac{\pi}{2} H_0^{(1)}(\sqrt{1-x_2} b_2) \right),
		\end{align}
		
		% 3. Annihilation A1-B-Pseudoscalar (th1r83 + th1r84)
		\begin{align}
\mathcal{M}^{P(b)-}_{A1_{k}} &= \int_0^1 dx_2 \int_0^{1/\Lambda} b_2 db_2 \, \eta_t \cdot \mathcal{C} Q_{q} 
			a_k(t_{A1}^b) \exp[-S_V(t_{A1}^b)] S_t(x_2)  \nonumber \\
			&\quad \times m_0 \left[ (x_2-2) \phi_V^a(x_2) - x_2 \phi_V^v(x_2) \right] 
			\left( -i \frac{\pi}{2} H_0^{(1)}(\sqrt{1-x_2} b_2) \right),
		\end{align}
		
		% 4. Annihilation A1-C (th1r85)
		\begin{align}
\mathcal{M}^{S(c)-}_{A1_{k}} &= \int_{x_1^d}^{x_1^u} dx_1 \int_0^{1/\Lambda} b_1 db_1 \int_0^\infty k_T dk_T \, \eta_t \cdot \mathcal{C} Q_{q'} a_k(t_{A1}^c) \exp[-S_B(t_{A1}^c)] S_t(x_1)\left[\right. 4 m_B m_0  \nonumber \\
			&\quad \times  (E_q + k_z)(E_Q + m_Q)\left. \right] 
			\left( \frac{m_B}{2} + \frac{k_T^2}{2 x_1^2 m_B} \right)
			K(k) J_0(k_T b_1) K_0(\sqrt{x_1} b_1),
		\end{align}
		
		\begin{align}
			\mathcal{M}^{P(c)-}_{A1_{k}} &= -\mathcal{M}^{S(c)-}_{A1_{k}},
		\end{align}
		
		% 5. Annihilation A1-D-Scalar (th1r86 + th1r87)
		\begin{align}
\mathcal{M}^{S(d)-}_{A1_{k}} &= \int_0^1 dx_2 \int_0^{1/\Lambda} b_2 db_2 \, \eta_t \cdot \mathcal{C} Q_{q''}
			a_k(t_{A1}^d) \exp[-S_V(t_{A1}^d)] S_t(x_2)  \nonumber \\
			&\quad \times m_0 \left[ (x_2-1)\phi_V^a(x_2) + (x_2+1)\phi_V^v(x_2) \right] 
			\left( -i \frac{\pi}{2} H_0^{(1)}(\sqrt{x_2} b_2) \right),
		\end{align}
		
		% 6. Annihilation A1-D-Pseudoscalar (th1r88 + th1r89)
		\begin{align}
\mathcal{M}^{P(d)-}_{A1_{k}} &= \int_0^1 dx_2 \int_0^{1/\Lambda} b_2 db_2 \, \eta_t \cdot \mathcal{C} Q_{q''} 
			a_k(t_{A1}^d) \exp[-S_V(t_{A1}^d)] S_t(x_2)  \nonumber \\
			&\quad \times (-m_0) \left[ (x_2+1)\phi_V^a(x_2) + (x_2-1)\phi_V^v(x_2) \right] 
			\left( -i \frac{\pi}{2} H_0^{(1)}(\sqrt{x_2} b_2) \right),
		\end{align}
	\end{widetext}

where $k$ in the subscript runs from 3 to 6. The diagrams in  Fig. \ref{fig-annihilation-b} only contribute to the neutral decay modes. Then the contributions of $(V-A)(V-A)$ type operators to the amplitudes of the decay modes $B^0 \to \rho^0 \gamma$ and $B^0 \to \omega \gamma$ are 	
\begin{equation}
	\begin{split}
		 &M(B^0 \to \rho^0 \gamma)^{j^-}_{A1} =\\
&\quad \frac{1}{\sqrt{2}} \bigg[ \left( M^{j(b)^-}_{A1_3} (Q_u)- M^{j(b)^-}_{A1_3} (Q_d) \right)  \\
& \quad + \left( M^{j(d)^-}_{A1_3} (Q_u) - M^{j(d)^-}_{A1_3} (Q_d) \right) \bigg],
	\end{split}
\end{equation}

\begin{equation}
	\begin{split}
		&M(B^0 \to \omega \gamma)^{j^-}_{A1} = \\
&\quad\frac{1}{\sqrt{2}} \bigg[  2M^{j(a)^-}_{A1_3} (Q_b) + \Big\{ M^{j(b)^-}_{A1_3} (Q_u) \\
& + M^{j(b)^-}_{A1_3} (Q_d) \Big\} + 2M^{j(c)^-}_{A1_3} (Q_d) \\
& + \Big\{ M^{j(d)^-}_{A1_3} (Q_u) + M^{j(d)^-}_{A1_3} (Q_d) \Big\} \bigg].
	\end{split}
\end{equation}

The contributions of operators of $(V-A)(V+A)$ type to the amplitudes are tagged by a symbol ``+" in the superscript, which are
\begin{equation}
	\begin{split}
		M_{A1}^{S(a)^+} (Q_b) &= M_{A1}^{S(a)^-} (Q_b), \\
		M_{A1}^{P(a)^+} (Q_b) &= M_{A1}^{P(a)^-} (Q_b), \\[6pt]
		M_{A1}^{S(b)^+} (Q_q) &= M_{A1}^{S(b)^-} (Q_q), \\
		M_{A1}^{P(b)^+} (Q_q) &= -M_{A1}^{P(b)^-} (Q_q), \\[6pt]
		M_{A1}^{S(c)^+} (Q_{q'}) &= M_{A1}^{S(c)^-} (Q_{q'}), \\
		M_{A1}^{P(c)^+} (Q_{q'}) &= M_{A1}^{P(c)^-} (Q_{q'}), \\[6pt]
		M_{A1}^{S(d)^+} (Q_{q''}) &= M_{A1}^{S(d)^-} (Q_{q''}), \\
		M_{A1}^{P(d)^+} (Q_{q''}) &= -M_{A1}^{P(d)^-} (Q_{q''}),
	\end{split}
\end{equation}
and the contributions of of $(V-A)(V+A)$ operators to the physical amplitudes are
\begin{equation}
	\begin{split}
&M(B^0 \to \rho^0 \gamma)^{j^+}_{A1} = \\
&\quad \frac{1}{\sqrt{2}} \bigg[  \Big\{ M^{j(b)^+}_{A1_5} (Q_u) - M^{j(b)^+}_{A1_5} (Q_d) \Big\} \\
&\quad + \Big\{ M^{j(d)^+}_{A1_5} (Q_u) - M^{j(d)^+}_{A1_5} (Q_d) \Big\} \bigg],
	\end{split}
\end{equation}

\begin{equation}
	\begin{split}
&M(B^0 \to \omega \gamma)^{j^+}_{A1} = \\
&\quad\frac{1}{\sqrt{2}} \bigg[  2M^{j(a)^+}_{A1_5} (Q_b) + \Big\{ M^{j(b)^+}_{A1_5} (Q_u) \\
&\quad + M^{j(b)^+}_{A1_5} (Q_d) \Big\} + 2M^{j(c)^+}_{A1_5} (Q_d) \\
&\quad + \Big\{ M^{j(d)^+}_{A1_5} (Q_u) + M^{j(d)^+}_{A1_5} (Q_d) \Big\} \bigg].
	\end{split}
\end{equation}

For the contributions of the annihilation diagrams shown in Fig. \ref{fig-annihilation-c} with $(V-A)(V-A)$ insertion, the results $M_{A2}^{(S,P)^{-}}$  are the same as $M_{A1}^{(S,P)^{-}}$, and the amplitudes for the decay modes are 
\begin{align}
	&M(B^{+} \to \rho^{+} \gamma)_{A2}^{j^{-}} = M_{A2_4}^{j(a)^{-}}(Q_{b})  \nonumber \\
	&\quad + M_{A2_4}^{j(b)^{-}}(Q_{d})+ M_{A2_4}^{j(c)^{-}}(Q_{u}) + M_{A2_4}^{j(d)^{-}}(Q_{u}), \\[10pt]
	 &M(B^{0} \to \rho^{0} \gamma)_{A2}^{j^{-}}= -\frac{1}{\sqrt{2}} \biggl[ M_{A2_4}^{j(a)^{-}}(Q_{b})  \nonumber \\
	&\qquad + M_{A2_4}^{j(b)^{-}}(Q_{d})+ M_{A2_4}^{j(c)^{-}}(Q_{d}) + M_{A2_4}^{j(d)^{-}}(Q_{d}) \biggr], \\[10pt]
	&M(B^{0} \to \omega \gamma)_{A2}^{j^{-}} = \frac{1}{\sqrt{2}} \biggl[ M_{A2_4}^{j(a)^{-}}(Q_{b})  \nonumber \\
	&\qquad + M_{A2_4}^{j(b)^{-}}(Q_{d})+ M_{A2_4}^{j(c)^{-}}(Q_{d}) + M_{A2_4}^{j(d)^{-}}(Q_{d}) \biggr].
\end{align}
	
While for the insertion of $(V-A)(V+A)$ operators, the results are
		\begin{align}
			\mathcal{M}^{S(a)+}_{A2_{k}} &= \mathcal{M}^{P(a)+}_{A2_{k}}=0,
		\end{align}
		% 1. Annihilation A2+B2 (th1r72 + th1r73)
		% 对应代码中的 c6 + c5/3 -> a6
		% ta = -4 * mB^2 * phit
		% 系数 1/(2*mB) * ta = -2 mB	

		\begin{align}
&\mathcal{M}^{S(b)+}_{A2_{k}} = \int_0^1 dx_2 \int_0^{1/\Lambda} b_2 db_2 \, \eta_t  \mathcal{C}_{ann}  Q_{d} 
			a_k(t_{A2}^b)  \nonumber \\
&\quad \times \exp[-S_V(t_{A2}^b)] S_t(x_2)\left[ -2 m_B \phi_V^T(x_2) \right] \nonumber \\
&\quad \times \left( -i \frac{\pi}{2} H_0^{(1)}(\sqrt{1-x_2} b_2) \right),
		\end{align}
		
		\begin{align}
			\mathcal{M}^{P(b)+}_{A2_{k}} &= -\mathcal{M}^{S(b)+}_{A2_{k}},
		\end{align}
		
		\begin{align}
			\mathcal{M}^{S(c)+}_{A2_{k}} &= \mathcal{M}^{P(c)+}_{A2_{k}}=0,
		\end{align}
		
		% 2. Annihilation A2+D2 (th1r74 + th1r75)
		% 对应代码中的 c6 + c5/3 -> a6
		% ta = 4 * mB^2 * phit
		% 系数 1/(2*mB) * ta = 2 mB
		\begin{align}
 &\mathcal{M}^{S(d)+}_{A2_{k}}= \int_0^1 dx_2 \int_0^{1/\Lambda} b_2 db_2 \, \eta_t  \mathcal{C}_{ann}  Q_{q'}
			a_k(t_{A2}^d) \nonumber \\
&\quad \times  \exp[-S_V(t_{A2}^d)] S_t(x_2)\left[ 2 m_B \phi_V^T(x_2) \right]  \nonumber \\
&\quad\left( -i \frac{\pi}{2} H_0^{(1)}(\sqrt{x_2} b_2) \right),
		\end{align}
		\begin{align}
			\mathcal{M}^{P(d)+}_{A2_{k}} &= -\mathcal{M}^{S(d)+}_{A2_{k}},
		\end{align}
where 
\begin{align*}
\mathcal{C}_{ann} = \frac{e m_B^2 f_B G_F}{\sqrt{27}}.
\end{align*}		
The amplitudes corresponding to the annihilation diagrams in Fig. \ref{fig-annihilation-c} with $(V-A)(V+A)$ operator insertion are
\begin{align}
 &M(B^+ \to \rho^+\gamma)_{A2}^{j^+}=M_{A2_6}^{j(a)^+}(Q_b)  \notag \\
	&\quad  + M_{A2_6}^{j(b)^+}(Q_d) + M_{A2_6}^{j(c)^+}(Q_u) + M_{A2_6}^{j(d)^+}(Q_u), \\[10pt]
&M(B^0 \to \rho^0\gamma)_{A2}^{j^+} = -\frac{1}{\sqrt{2}} \Bigl[ M_{A2_6}^{j(a)^+}(Q_b)  \notag \\
	&\quad + M_{A2_6}^{j(b)^+}(Q_d)+ M_{A2_6}^{j(c)^+}(Q_d) + M_{A2_6}^{j(d)^+}(Q_d) \Bigr], \\[10pt]
 &M(B^0 \to \omega\gamma)_{A2}^{j^+}= \frac{1}{\sqrt{2}} \Bigl[ M_{A2_6}^{j(a)^+}(Q_b)  \notag \\
	&\quad + M_{A2_6}^{j(b)^+}(Q_d)+ M_{A2_6}^{j(c)^+}(Q_d) + M_{A2_6}^{j(d)^+}(Q_d) \Bigr].
\end{align}

Finally the total amplitudes of all the decay modes for $j=S,P$ are
\begin{widetext}
	\begin{align}
&M^{j}(B^{+} \rightarrow \rho^{+} \gamma) = M(B^{+} \rightarrow \rho^{+} \gamma)_{7\gamma}^{j} + M(B^{+} \rightarrow \rho^{+} \gamma)_{8g}^{j} \notag + M(B^{+} \rightarrow \rho^{+} \gamma)_{1i}^{j} \notag \\
	&\qquad + M(B^{+} \rightarrow \rho^{+} \gamma)_{2i}^{j} + M(B^{+} \rightarrow \rho^{+} \gamma)_{A}^{j} + M(B^{+} \rightarrow \rho^{+} \gamma)_{A2}^{j-}  + M(B^{+} \rightarrow \rho^{+} \gamma)_{A2}^{j+}, \\[12pt]
&M^{j}(B^{0} \rightarrow \rho^{0} \gamma) = M(B^{0} \rightarrow \rho^{0} \gamma)_{7\gamma}^{j} + M(B^{0} \rightarrow \rho^{0} \gamma)_{8g}^{j} \notag + M(B^{0} \rightarrow \rho^{0} \gamma)_{1i}^{j} + M(B^{0} \rightarrow \rho^{0} \gamma)_{2i}^{j} \notag \\
	&\qquad + M(B^{0} \rightarrow \rho^{0} \gamma)_{A}^{j} + M(B^{0} \rightarrow \rho^{0} \gamma)_{A1}^{j-}  + M(B^{0} \rightarrow \rho^{0} \gamma)_{A1}^{j+} + M(B^{0} \rightarrow \rho^{0} \gamma)_{A2}^{j-} \notag \\
	&\qquad + M(B^{0} \rightarrow \rho^{0} \gamma)_{A2}^{j+}, \\[12pt]
 &M^{j}(B^{0} \rightarrow \omega \gamma)= M(B^{0} \rightarrow \omega \gamma)_{7\gamma}^{j} + M(B^{0} \rightarrow \omega \gamma)_{8g}^{j} + M(B^{0} \rightarrow \omega \gamma)_{1i}^{j} + M(B^{0} \rightarrow \omega \gamma)_{2i}^{j} \notag \\
	&\qquad + M(B^{0} \rightarrow \omega \gamma)_{A}^{j} + M(B^{0} \rightarrow \omega \gamma)_{A1}^{j-} \quad + M(B^{0} \rightarrow \omega \gamma)_{A1}^{j+} + M(B^{0} \rightarrow \omega \gamma)_{A2}^{j-} \notag \\
	&\qquad + M(B^{0} \rightarrow \omega \gamma)_{A2}^{j+}.
	\end{align}
\end{widetext}

\section{ THE CONTRIBUTION OF SOFT TRANSITION FORM FACTORS}

Only when the scale for the hard function $H$ is high enough, can the factorization formula given in Eq. (\ref{eq-factorization}) holds. Therefore, the perturbative calculations in PQCD approach explained in Secs. \ref{sec-hard}, \ref{sec-NLO} and \ref{sec-annihilation} must be restricted in the regions with the scale $\mu>\mu_c$, where $\mu_c$ is the infrared momentum cutoff scale which serves to separate the hard and soft interactions. It is reasonable to choose $\mu_c$ around 1 GeV  in QCD. For the interactions in the hard regime  $\mu=t_i>\mu_c$, the contributions to the decay amplitudes are calculated with PQCD approach, here $t_i$'s stand for the hard scales chosen for each diagram relevant to the hard transition process. While for the contributions with the scale $\mu<\mu_c$, we can introduce soft form factors to absorb the soft contribution.

%\begin{widetext}
%\begin{equation}
%	\begin{aligned}
%		\langle V(P_1) | \bar{q}  \sigma_{\mu\nu} b | \bar{B}(P_B) \rangle  &= -i \varepsilon_{\mu\nu\alpha\beta} %\epsilon_V^{\alpha *} P_1^\beta T_1(q^2) - i \varepsilon_{\mu\nu\alpha\beta} \epsilon_V^{\alpha *} P_B^\beta T_2(q^2) \\
%		&\quad - i T_3(q^2) \frac{P_B \cdot \epsilon_V^*}{P_B \cdot P_1} \varepsilon_{\mu\nu\alpha\beta} P_1^\alpha %P_B^\beta
%	\end{aligned}
%\end{equation}
%\begin{equation}
%	\begin{aligned}
%		\langle V(P_1) | \bar{q}  \sigma^{\mu\nu} \gamma_5 b | \bar{B}(P_B) \rangle  &= (P_1^\mu P_B^\nu - P_B^\mu %P_1^\nu) \frac{P_B \cdot \epsilon_V^*}{P_B \cdot P_1} T_3(q^2) \\
%		&\quad + (\epsilon_V^{\mu *} P_B^\nu - P_B^\mu \epsilon_V^{\nu *}) T_2(q^2) + (\epsilon_V^{\mu *} P_1^\nu - %P_1^\mu \epsilon_V^{\nu *}) T_1(q^2)
%	\end{aligned}
%\end{equation}
%\end{widetext}
%\begin{equation}
%\begin{aligned}
%	&\langle V(p_1) | \bar{q} \sigma_{\mu\nu} q^\nu (1+\gamma_5) b | \bar{B}(p_B) \rangle \\
%	&= 2i \varepsilon_{\mu\nu\alpha\beta} \epsilon_V^{\nu*} p_B^\alpha p_1^\beta T_1'(q^2) \\
%	&\quad + \left[ \epsilon_{V\mu}^* (m_B^2 - m_V^2) - (q \cdot \epsilon_V^*) (p_1 + p_B)_\mu \right] T_2'(q^2) \\
%	&\quad + (q \cdot \epsilon_V^*) \left[ q_\mu - \frac{q^2}{m_B^2 - m_V^2} (p_1 + p_B)_\mu \right] T_3'(q^2)
%\end{aligned}
%\end{equation}

The transition form factors relevant to the radiative decays of $B$ meson can be defined by the matrix element of the tensor current \cite{Col-1996}
\begin{equation}\label{fm-Ti}
\begin{aligned}
	&\langle V(p_1) | \bar{q} \sigma_{\mu\nu} q^\nu (1+\gamma_5) b | \bar{B}(p_B) \rangle \\
	&= 2i \varepsilon_{\mu\nu\alpha\beta} \epsilon_V^{\nu*} p_B^\alpha p_1^\beta T_1(q^2)
       + \left[\right. \epsilon_{V\mu}^* (m_B^2 - m_V^2)  \\
	&\quad - (q \cdot \epsilon_V^*) (p_1 + p_B)_\mu\left. \right] T_2(q^2) + (q \cdot \epsilon_V^*)\\
	&\quad \times \left[ q_\mu - \frac{q^2}{m_B^2 - m_V^2} (p_1 + p_B)_\mu \right] T_3(q^2),
\end{aligned}
\end{equation}
%\begin{equation}
%\begin{aligned}
%	T_1'(q^2) &= \frac{1}{2} \left[ T_1(q^2) + T_2(q^2) \right] \\
%	T_2'(q^2) &= \frac{1}{2} \left[ T_1(q^2) + T_2(q^2) + \frac{q^2}{m_B^2 - m_V^2} (T_2(q^2) - T_1(q^2)) \right] \\
%	T_3'(q^2) &= \frac{1}{2} \left[ T_1(q^2) - T_2(q^2) - \frac{m_B^2 - m_V^2}{p_B \cdot p_1} T_3(q^2) \right]
%\end{aligned}
%\end{equation}
where $T_{1,2,3}(q^2)$ are the transition form factors corresponding to the tensor current, and there is the relation $T_1(0) = T_2(0)$. With the form factors defined by the tensor current, the matrix element induced by the operator $O_{7\gamma}$ can be derived as 
%\begin{widetext}
\begin{equation} \label{amplitude-O7gamma}
\begin{aligned}
&\langle V \gamma | Q_{7\gamma} | \bar{B}(p_B) \rangle= \frac{ie}{4\pi^2} m_b \bigg[ (\epsilon_\gamma^* \cdot \epsilon_V^*) (m_B^2 - m_V^2) \\
	&\qquad \times T_2(0)+ 2i \varepsilon_{\mu\nu\alpha\beta} \epsilon_\gamma^{*\mu} \epsilon_V^{*\nu} p_\gamma^\alpha p_V^\beta T_1(0) \bigg].
\end{aligned}
\end{equation}
%\end{widetext}

There is actually only one independent form factor that contributes, because of the relation $T_1(0)=T_2(0)$. The form factors for $B\to\rho$ and $B\to\omega$ can be separated into hard and soft form factors as the follows
\begin{equation} \label{hardsoft}
\begin{aligned}
	T_{1\rho}(0) &= h_{\rho} + \xi_{\rho},\\
%\end{equation}
%\begin{equation}\label{harsoft-b)
T_{1\omega}(0) &= h_{\omega} + \xi_{\omega},
\end{aligned}
\end{equation}
where $h_{\rho,\omega}$ denote the hard form factors, which can be calculated with PQCD approach. The hard contributions to the decay amplitudes have been given in Sec. \ref{sec-hard}, which are corresponding to the diagrams in Fig. \ref{fig-O7gamma}. $\xi_{\rho,\omega}$ are the soft form factors, which are the nonperturbative quantities and absorb the soft contributions. The soft form factors are treated as input parameters in this work.

One can obtain the contributions of the soft form factors to $M^{S,P}$ by using the definition of $M^{S,P}$ in Eq. (\ref{MSMP}) and Eqs. (\ref{amplitude-O7gamma}) and (\ref{hardsoft}) by only considering  soft part
\begin{equation}
\begin{aligned}
	M^S_{\mathrm{soft}} &= \frac{G_F}{\sqrt{2}} (-V_{tb}^* V_{td})\frac{e}{4\pi^2} m_b (m_B^2 - m_V^2)C_{7\gamma}(\mu_c) \\
	&\quad\times\xi_{\rho,\omega}(0),  \\
	M^P_{\mathrm{soft}} &=  \frac{G_F}{\sqrt{2}} (-V_{tb}^* V_{td})\frac{e}{2\pi^2} m_b p_V \cdot p_\gamma C_{7\gamma}(\mu_c)\\
	&\quad\times\xi_{\rho,\omega}(0).
\end{aligned}
\end{equation}
Then the amplitude $M^{S,P}$ calculated by PQCD should be replaced by 
      \begin{equation}
      M^{S,P}\to  M^{S,P}+M^{S,P}_{\mathrm{soft}}.
      \end{equation}

\section{NUMERICAL RESULT AND DISCUSSION}

For the numerical treatment, the input parameters used in this work are collected in Table \ref{tab-inputs}. The CKM parameters in Wolfenstein parameterization are $\lambda$, $A$, $\rho$ and $\eta$. The definition $\bar{\rho}=\rho(1-\lambda^2/2)$ and $\bar{\eta}=\eta(1-\lambda^2/2)$ are usually used. The CKM parameters, meson masses and life times can be found in PDG \cite{PDG2024}. The decay constant $f_B$ is the result calculated in the relativistic potential model from Ref. \cite{SY2019},  and the other decay constants of the light vector mesons are kept the same value as Ref. \cite{Lu-etal2005}.

% 删掉这里的 \begin{widetext} 和 \begin{center}
		
		\begin{table*}[htbp] % 注意这里的星号 *
			\caption{Summary of the input parameters.}\label{tab-inputs}
			\begin{ruledtabular}
				\begin{tabular}{ccccc}
					\multicolumn{5}{c}{CKM parameters and QCD constant} \\
					$\lambda$ & $A$ & $\bar{\rho}$ & $\bar{\eta}$ & $\Lambda^{(f=4)}_{\overline{\text{MS}}}$ \\
					$0.22501\pm 0.00068$ & $0.826^{0.016}_{-0.015}$ & $0.1591 \pm 0.0094$ & $0.3523^{+0.0073}_{-0.0071}$ & 250 MeV \\
					\hline
					
					\multicolumn{5}{c}{Masses} \\
					$M_W$ & $M_B$ & $M_\rho$ & $M_\omega$ & $m_c$ \\
					80.41 GeV & 5.28 GeV & 0.78 GeV & 0.78 GeV & 1.5 MeV \\
					\hline
					\multicolumn{5}{c}{Meson decay constants} \\
					$f_B$ & $f_\rho$ & $f_\rho^T$ & $f_\omega$ & $f_\omega^T$ \\
					210 MeV & 220 MeV & 160 MeV & 195 MeV & 160 MeV \\
					\hline		
					\multicolumn{2}{c}{$B$ meson lifetime}& \multicolumn{2}{c}{$\mu_c $}&\\
					$\tau_{B^0}$ & $\tau_{B^\pm}$  &   & &   \\
					1.517 ps   &   1.638 ps      &\multicolumn{2}{c}{1 GeV} & \\
				\end{tabular}
			\end{ruledtabular}
		\end{table*} % 注意这里的星号 *
		
		% 删掉这里的 \end{center} 和 \end{widetext}
The decay width and branching ratio are calculated according to the following formulas
\begin{equation}
	\Gamma = \frac{|M^S|^2 + |M^P|^2}{8 \pi M_B},\qquad Br = \frac{\tau_B}{\hbar} \Gamma,
\end{equation}
and the branching ratios calculated in this work are defined as the average of the decay mode and its charge conjugate
\begin{equation}
\begin{split}
	Br(B^\pm \to \rho^\pm \gamma) &= \frac{1}{2} \left[\right. Br(B^+ \to \rho^+ \gamma)\\
	 &\quad + Br(B^- \to \rho^- \gamma)\left. \right],
	 \end{split}
\end{equation}

\begin{equation}
\begin{split}
	Br(B^0 \to \rho^0 \gamma) &= \frac{1}{2} \left[\right. Br(B^0 \to \rho^0 \gamma) \\
	&\quad+ Br(\bar{B}^0 \to \rho^0 \gamma) \left.\right].
	 \end{split}
\end{equation}

The direct $CP$ asymmetry parameters for the charged and neutral decay modes are defined by
\begin{equation}
\begin{split}
	&A_{cp}(B^\pm \to \rho^\pm \gamma) \\
	&= \frac{\Gamma(B^- \to \rho^- \gamma) - \Gamma(B^+ \to \rho^+ \gamma)}{\Gamma(B^- \to \rho^- \gamma) + \Gamma(B^+ \to \rho^+ \gamma)},
\end{split}
\end{equation}
and
\begin{equation}
\begin{split}
	&A_{cp}(B^0 \to \rho^0(\omega)\gamma) \\
	&= \frac{\Gamma(\bar{B}^0 \to \rho^0(\omega)\gamma) - \Gamma(B^0 \to \rho^0(\omega)\gamma)}{\Gamma(\bar{B}^0 \to \rho^0(\omega)\gamma) + \Gamma(B^0 \to \rho^0(\omega)\gamma)}.\\
	& \\
	&
\end{split}
\end{equation}	
 Using the input parameters given in Table \ref{tab-inputs}, we calculate the branching ratios and $CP$ violations of the decays $B\to \rho(\omega)\gamma$ with the soft form factors as unknown parameters, and confront these quantities to experimental data. By fitting the experimental data for the branching ratios and $CP$ violation, we obtain the values of the soft form factors, which are shown in Table \ref{value-fm}. The errors mainly come from the uncertainties of the parameters in wave functions of $B$ meson, light mesons and the errors of the experimental data. We also calculate the hard form factors $h_\rho$ and $h_\omega$ in PQCD by restricting the momentum scale to be $\mu>\mu_c$, which are also shown in Table \ref{value-fm}.
The total form factors $T_{1\rho}(0)$ and $T_{1\omega}(0)$ can be obtained by summing the the hard and soft form factors. The comparison of the total form factors obtained in this work and that from LCSR \cite{Ball-etal2005} can be found in the last two rows in Table \ref{value-fm}, which shows the consistence of the fitted results of the soft form factors in the modified PQCD approach and the form factors calculated from LCSR method.
\begin{table}[htbp]
	\centering
	\caption{ The form factors }\label{value-fm}
	\begin{tabular}{cccc}
		\hline\hline
		& PQCD & LCSR \\
		\hline
		$\xi_{\rho}$ &$0.10^{+0.02}_{-0.02}$  &- \\
		$\xi_{\omega}$ &$0.07^{+0.01}_{-0.02}$  &- \\
		$h_{\rho}$ &$0.18^{+0.01}_{-0.01}$  &- \\
		$h_{\omega}$ &$0.17^{+0.01}_{-0.01}$  &- \\
		$T_{1\rho}(0)$ &$0.28^{+0.03}_{-0.03}$  &$0.267\pm 0.021$ \\
		$T_{1\omega}(0)$ &$0.24^{+0.03}_{-0.03}$  &$0.242\pm 0.022$ \\
		\hline\hline
	\end{tabular}
\end{table}

With the soft form factors presented in Table \ref{value-fm}, the amplitudes $M^S$ and $M^P$ for each decay mode can be calculated, which are shown in Table \ref{tab-amplitude}. It is shown that the contributions of $O_{7\gamma}$ are overwhelmingly dominant in all the contributions. The NLO contribution from quark loop and annihilation diagrams in $B^+\to\rho^{+}\gamma$ is larger than that in other decay modes. The column ``All" denotes the contributions of all the leading and next-to-leading order diagrams except the contribution of the soft form factor. The column ``$\xi_{\mathrm{soft}}$" means the contributions of the soft form factor. This table shows that the relative contribution of the soft form factor respect to hard contribution in the neutral decay mode is larger than that in the charged mode. 

The theoretically calculated results of the branching ratios and $CP$ violations for the decay modes considered in this work are shown in Table \ref{value-brcp}, where ``LO" denotes the leading-order contributions corresponding to the operators $O_{7\gamma}$ and $O_{8g}$, ``+NLO" the results including the contributions up to the next-to-leading order diagrams (the quark-loop and annihilation diagrams), and ``+NLO+$\xi_V$" the total contributions including both the LO, NLO and the soft form factor contributions. The comparison of LO and NLO shows that the differences of these two columns are very small, which means the the contributions of NLO diagrams are not large. The difference is smaller than a few percent, which indicates that the perturbative calculation is indeed effective in the modified PQCD approach. The contribution of the soft form factor slightly lower the branching ratio of $B^{+}\to\rho^{+}\gamma$ decay mode, and affect the $CP$ violation largely. With the soft form factors given in Table \ref{value-fm}, the theoretical outputs of both the branching ratios and $CP$ violations are well consistent with the experimental data.
\begin{widetext}
\begin{center}
	\begin{table}[htbp]
		\centering
		\caption{The amplitudes of individual Feynman diagrams and soft contributions  $( \times 10^{-9})$.}
\label{tab-amplitude}
		\begin{tabular}{cccccccc}
			\toprule
	& & decay mode		& $O7\gamma$                                         & O8g    & NLO             & ALL &$\xi_{\mathrm{soft}}$ \\
			\midrule
	&\vline	&	$B^+\to\rho^{+}\gamma$    &$5.00+2.40i$   &$ 0.08+0.17i$   & $0.67-2.86i$       & $5.76-0.27i$ &$-1.56+3.22i$ \\
	$M^{S}$ &\vline	&	$B^0\to\rho^{0}\gamma$    &$-3.54-1.71i$   &$ -0.12-0.02i$   & $-0.26-0.23i$       & $-3.91-1.96i$ &$1.10-2.28i$ \\
		&\vline&	$B^0\to\omega\gamma$  &$3.27+1.58i$   &$ 0.12+0.02i$   & $-0.20+0.03i$       & $3.19+1.63i$ &$-1.10+2.28i$ \\ \hline
		&\vline&	$B^+\to\rho^{+}\gamma$    &$-5.00-2.42i$   &$ -0.10-0.17i$   & $-1.84+2.34i$       & $-6.94-0.25i$ &$1.56-3.22i$ \\
	$M^{P}$&\vline	&	$B^0\to\rho^{0}\gamma$    &$3.54+1.74i$   &$ 0.11+0.02i$   & $-0.03-0.03i$       & $3.62+1.70i$ &$-1.10+2.28i$ \\
		&\vline&	$B^0\to\omega\gamma$  &$-3.27-1.58i$   &$ -0.11-0.02i$   & $0.04+0.03i$       & $-3.34-1.57i$ &$1.10-2.28i$ \\
			\bottomrule
		\end{tabular}
	\end{table}
\end{center}
\end{widetext}		
\begin{table*}[htbp] % 使用 table* 可以让表格跨双栏显示（如果在双栏排版中表格太宽）
	\caption{\label{tab:results_double} Branching ratios and $CP$ asymmetries with asymmetric errors (Double-line display).}\label{value-brcp}
	\renewcommand{\arraystretch}{1.8} % 进一步增加行高，因为有上下标，行距需要更大
	\setlength{\tabcolsep}{6pt}       % 适当调整列间距
	
	\begin{tabular}{ccccc}
		\hline\hline
		% 第一行表头：使用 \shortstack 强制换行，或者用 \substack (需在数学模式下)
		Mode & LO & +NLO &  +NLO$+$$\xi_{V}$  & Data \cite{PDG2024} \\ 
		\hline
		Br($B^{+}\to\rho^{+}\gamma$)$\times$ $10^{-6}$   
		& 1.31 
		& 1.38  % 双行误差写法
		& $1.34^{+0.12}_{-0.05}$ 
		&$1.29 \pm 0.20$ \\
		
		Br($B^{0}\to\rho^{0}\gamma$)$\times$ $10^{-6}$   
		& 0.56  
		& 0.58 
		& $0.81^{+0.02}_{-0.01}$
		& $0.82 \pm 0.13$ \\
		
		Br($B^{0}\to\omega\gamma$)$\times$ $10^{-6}$  
		& 0.48  
		& 0.50 
		& $0.60^{+0.13}_{-0.23}$ 
		& $0.44 \pm 0.17$ \\
		\hline
		$A_{CP}(B^{+}\to\rho^{+}\gamma)$     
		& 0.127  
		& 0.108 
		& $-0.058^{+0.031}_{-0.022}$ 
		& $-0.08 \pm 0.15$ \\
		
		$A_{CP}(B^{0}\to\rho^{0}\gamma)$     
		& 0.064   
		& 0.045  
		& $0.037^{+0.001}_{-0.001}$  
		& - \\
		
		$A_{CP}(B^{0}\to\omega\gamma)$    
		& -0.062  
		& -0.083 
		& $-0.063^{+0.009}_{-0.026}$ 
		& - \\
		\hline\hline
	\end{tabular}
\end{table*}
%--- 表格代码结束 ---		

\section{SEEKING MASSLESS DARK PHOTONS IN THE DECAYS OF BOTTOM MESON}

In the SM electromagnetic interaction is described by the $\mathrm{U(1)_{em}}$ symmetry group, which is the consequence of the spontaneous breaking of the $\mathrm{SU}(2)_L\times \mathrm{U}(1)_Y$ symmetry, where $L$ denotes the left-handed states of the fermion doublets, and $Y$ the hypercharge of the particles. In principle, SM can be extended by introducing an extra U(1) symmetry group \cite{Holdom-1986,Chiang-2017,He-2018}, which governs a sector of matter of darkness \cite{Gab-2014}. The symmetry of the extra U(1) group can be unbroken, then a massless U(1) gauge boson will be generated, which can be called the dark photon \cite{Dob-2005,Gab-2016}. The only connection between the dark sector and the matter particles in SM is through the dark photon. The effective Lagrangian describing the interaction between the SM fermion and the massless dark photon is \cite{Dob-2005,Su-2020} 

\begin{equation}
\begin{split}
	\mathcal{L}_{\mathrm{NP}}& = \frac{1}{\Lambda_{\mathrm{NP}}^2} (\mathcal{C}_{jk}^{D} \overline{q}_j \sigma^{\mu\nu} d_k H + \mathcal{C}_{jk}^{U} \overline{q}_j \sigma^{\mu\nu} u_k \tilde{H}\\
&\quad + \mathrm{H.c.}) \bar{F}_{\mu\nu},
\end{split}
\end{equation}
where $\Lambda_{\mathrm{NP}}$ denotes the large mass scale for the new physics of dark sector, $\mathcal{C}_{jk}^{U,D}$'s the couplings of the quark fields and the Higgs scalar, $\overline{q}_j$ the left-handed quark doublet of SU(2)$_L$ group, $d_k$ ($u_k$) the right-handed down (up)-type quark singlet, $H$ the SM Higgs doublet, and $\tilde{H}=i\sigma_2 H^*$, where $\sigma_2$ is the second Pauli metrix. $\bar{F}_{\mu\nu}=\partial_\mu \bar{A}_\nu-\partial_\nu \bar{A}_\mu$ is the dark photon strength tensor. The indices $i,j,k=1,2,3$.

After electroweak symmetry breaking, the terms for the $b\to d$ FCNC current interaction with the massless dark photon can be extracted from the  $\mathcal{C}_{jk}^{D}$ sector
\begin{equation}\label{effective-L-b}
\mathcal{L}_{db\gamma^\prime} = \bar{d}(\mathbb{C} + \gamma_5 \mathbb{C}_5)\sigma^{\mu\nu} b \bar{F}_{\mu\nu} + \mathrm{H.c.},
\end{equation}
where
\begin{equation}
\mathcal{C}=\Lambda_{\mathrm{NP}}^{-2}(\mathcal{C}_{13}^{D}+\mathcal{C}_{31}^{D*})v/\sqrt{8}
\end{equation}
and  
\begin{equation}
\mathbb{C}_5=\Lambda_{\mathrm{NP}}^{-2}(\mathcal{C}_{13}^{D}-\mathcal{C}_{31}^{D*})v/\sqrt{8}.
\end{equation}

Using the effective Lagrangian presented in Eq. (\ref{effective-L-b}) to calculate the decay amplitude, the matrix elements $\langle V(P_1) | \bar{q} \sigma_{\mu\nu} b | \bar{B}(P_B) \rangle $ and  $\langle V(P_1) | \bar{q} \sigma_{\mu\nu} \gamma_5 b | \bar{B}(P_B) \rangle$ should be treated. Three form factors can be defined for these two matrix elements
\begin{widetext}
\begin{equation}
\begin{aligned}
	\langle V(P_1) | \bar{q} \sigma_{\mu\nu} b | \bar{B}(P_B) \rangle  &=( -i \varepsilon_{\mu\nu\alpha\beta} \epsilon_V^{\alpha *} P_1^\beta \bar{T}_1(q^2) - i \varepsilon_{\mu\nu\alpha\beta} \epsilon_V^{\alpha *} P_B^\beta \bar{T}_2(q^2) - i \bar{T}_3(q^2) \frac{P_B \cdot \epsilon_V^*}{P_B \cdot P_1}\\
	&\quad \cdot \varepsilon_{\mu\nu\alpha\beta} P_1^\alpha P_B^\beta ),
\end{aligned}
\end{equation}
\begin{equation}
\begin{aligned}
	\langle V(P_1) | \bar{q} \sigma_{\mu\nu} \gamma_5 b | \bar{B}(P_B) \rangle  &= ((P_1^\mu P_B^\nu - P_B^\mu P_1^\nu) \frac{P_B \cdot \epsilon_V^*}{P_B \cdot P_1} \bar{T}_3(q^2) + (\epsilon_V^{\mu *} P_B^\nu - P_B^\mu \epsilon_V^{\nu *}) \bar{T}_2(q^2)\\
	&\quad  + (\epsilon_V^{\mu *} P_1^\nu - P_1^\mu \epsilon_V^{\nu *}) \bar{T}_1(q^2)),
\end{aligned}
\end{equation}
here $q^\nu=P_B^\nu-P_1^\nu$. Then for $\langle V(P_1) | \bar{q} \sigma_{\mu\nu}   q^\nu b | \bar{B}(P_B) \rangle$ and $\langle V(P_1) | \bar{q} \sigma^{\mu\nu} q_\nu \gamma_5 b | \bar{B}(P_B) \rangle$,  we have

\begin{equation}
\begin{aligned}
	 &\langle V(P_1) | \bar{q} \sigma_{\mu\nu} q^\nu b | \bar{B}(P_B) \rangle = 2 i \varepsilon_{\mu\nu\alpha\beta}\epsilon_{\gamma^\prime}^{\mu *} \epsilon_V^\nu P_B^{\alpha *} P_1^\beta T_1(q^2),\\
	 &\langle V(P_1) | \bar{q} \sigma^{\mu\nu}  q_\nu\gamma_5 b | \bar{B}(P_B) \rangle  =  \left[ \epsilon_{V\mu}^* (m_B^2 - m_V^2) - (q \cdot \epsilon_V^*) (p_1 + p_B)_\mu \right] T_2(q^2) \\
	&\quad + (q \cdot \epsilon_V^*) \left[ q_\mu - \frac{q^2}{m_B^2 - m_V^2} (p_1 + p_B)_\mu \right] T_3(q^2),
\end{aligned}
\end{equation}
\end{widetext}
where the relations between $T_i(q^2)$ and $\bar{T}_i(q^2)$ are 
\begin{equation}
\begin{aligned}
	T_1(q^2) &= \frac{1}{2} \left[ \bar{T}_1(q^2) + \bar{T}_2(q^2) \right], \\
	T_2(q^2) &= \frac{1}{2} \left[ \right.\bar{T}_1(q^2) + \bar{T}_2(q^2) + \frac{q^2}{m_B^2 - m_V^2} (\bar{T}_2(q^2)\notag\\
          &\quad - \bar{T}_1(q^2))\left. \right],\\
	T_3(q^2) &= \frac{1}{2} \left[ \bar{T}_1(q^2) -\bar{T}_2(q^2) - \frac{m_B^2 - m_V^2}{p_B \cdot p_1} \bar{T}_3(q^2) \right],
\end{aligned}
\end{equation}
Note that the form factors $T_i(q^2)$'s are the same as that in Eq. (\ref{fm-Ti}).
For the massless dark photon we have $\epsilon_{\gamma^\prime}^{*}\cdot q=0$ and $q^2=0$, then the amplitudes $M^S$ and $M^P$ for $B \to V \gamma^\prime$ decay can be calculated to be
\begin{equation}
\begin{aligned}
	\mathcal{M'}^S &= 2 (m_B^2 - m_V^2) T_2(0)\mathbb{C}_5,  \\
	\mathcal{M'}^P &= 4 P_V \cdot P_\gamma T_1(0)\mathbb{C}, 
\end{aligned},
\end{equation}
and the branching ratio is
\begin{equation}\label{br-dark-photon}
\begin{aligned}
\mathcal{B}r(B \to V \gamma^\prime) &= \frac{\tau_{B} T_{1}(0)^2 (m_{B}^2 - m_{V}^2)^2}{2\pi m_{B}} (|\mathbb{C}|^2 \\
   &+ |\mathbb{C}_5|^2).
\end{aligned}
\end{equation}
The form factor $T_{1}(0)$ is obtained in this work, and presented in Table \ref{value-fm}. For the parameters $\mathbb{C}$ and $\mathbb{C}_5$, we take
\begin{equation}\label{c-limit}
|\mathbb{C}|^2 + |\mathbb{C}_5|^2 < \frac{3.6 \times 10^{-18}}{\mathrm{GeV}^2},
\end{equation}
which is obtained by using the branching ratio $Br(b\to q \gamma^\prime)<1.0\times 10^{-5}$ (See Fig. 5 of Ref. \cite{Gab-2016}), and using Eq. (61) of \cite{Gab-2016}: 
\begin{equation}
\begin{split}
&Br(b\to q \gamma')=12Br^{\mathrm{exp}}(B\to X_c\bar{\nu}e)/[G^2_F|V_{cb}|^2m_b^2\\
&\quad \times f_1(m_c^2/m_b^2)]\cdot(1/(\Lambda_L^{bq})^2+1/(\Lambda_R^{bq})^2)
\end{split}
\end{equation}
with $B^{\mathrm{exp}}(Br\to X_c\bar{\nu}e)=0.108$ listed in PDG \cite{PDG2024}, $f_1(x)=1-8x+8x^3-x^4-12x^2\log x$ and utilizing the relation $1/(\Lambda_L^{bq})^2+1/(\Lambda_R^{bq})^2=8|\mathbb{C}|^2 +8 |\mathbb{C}_5|^2$.

With Eq. (\ref{br-dark-photon}), the upper limit in (\ref{c-limit}), and using the transition from factors $T_{1\rho}$, $T_{1\omega}$ presented in Table \ref{value-fm} for $T_1(0)$, we arrive at
\begin{equation}
\begin{split}
           &Br(B^{+}\to\rho^{+}\gamma')\textless1.55 \times 10^{-5},      \\
		&Br(B^{0}\to\rho^{0}\gamma')\textless1.44\times 10^{-5},       \\
		&Br(B^{0}\to\omega\gamma')\textless1.06\times 10^{-5}.   
\end{split}
\end{equation}
It is interesting for experiments of Belle II and LHCb to perform the corresponding detections, which can set constraint on the combination of the new physics parameters $|\mathbb{C}|^2 + |\mathbb{C}_5|^2$ on the experimental side.
	%\begin{table}[htbp]
	%\centering
	%\caption{$B\to\rho(\omega)\gamma'$ branching ratios.}
	%\begin{tabular}{cc}
	%	\toprule
	%	& $BR^{max}$ \\
	%	\midrule
	%	Br($B^{+}\to\rho^{+}\gamma'$)$\times$ $10^{-5}$&\textless1.55       \\
	%	Br($B^{0}\to\rho^{0}\gamma'$)$\times$ $10^{-5}$&\textless1.44       \\
	%	Br($B^{0}\to\omega\gamma'$)$\times$ $10^{-5}$  &\textless1.06      \\
	%	\bottomrule
	%\end{tabular}
     %\end{table}

\section{SUMMARY}
	\label{sec:conclusion}
We study the radiative decays of $B\to\rho\gamma$ and  $B\to\omega\gamma$ in the modified PQCD approach. The branching ratios and $CP$ violations are calculated with the transverse momenta of quark and gluons included. Sudakov factor is also included in the calculation, which helps to suppress the long-distance contributions. An infrared momentum cutoff scale $\mu_c$ is introduced to separate the soft and hard contributions. For the contributions with the interaction scale $\mu >\mu_c$, the amplitude can be calculated in the perturbative method. While, for the contributions with the scale $\mu <\mu_c$, soft form factors are introduced to collect the soft contributions. By fitting to the experimental data, we obtain the reasonable values for the soft form factors. The sum of the soft and hard form factors is consistent with the total form factor calculated with LCSR method. We also estimate the upper limit of the branching ratios of $B\to \rho$ and $B\to\omega$ transitions with the radiation of the massless dark photon by using the transition form factors obtained in the modified PQCD approach. The upper limits obtained in this work can be helpful  for detecting the existence of dark photon in $B$ decays.

	% -------------------------------------------------------------------
	% 致谢 (Acknowledgments)
	% -------------------------------------------------------------------
	\begin{acknowledgments}
This work is supported in part by the National Natural Science Foundation of China under Contracts No. 12275139, 12535006, 11875168.
	\end{acknowledgments}
	
	% -------------------------------------------------------------------
	% 附录 (可选)
	% -------------------------------------------------------------------
	\appendix
\section{THRESHOLD AND SUDAKOV FACTOR}\label{Appendix A}

The functions $S_B$ and $S_V$ in Eqs. (\ref{M7gamma-a}), (\ref{M7gamma-b}), etc. are given in the following 

\begin{equation}
	S_B(\mu) = s(x, b, m_B) - \frac{1}{\beta_1} \ln \frac{\ln(\mu / \Lambda_{\text{QCD}})}{\ln(1 / (b \Lambda_{\text{QCD}}))},
\end{equation}

\begin{eqnarray}
	S_{V}(\mu) &= s(x, b, m_B) + s(1 - x, b, m_B) \nonumber\\
 &- \frac{1}{\beta_1} \ln \frac{\ln(\mu / \Lambda_{\text{QCD}})}{\ln(1 / (b \Lambda_{\text{QCD}}))},
\end{eqnarray}
where the exponential $\exp{-s(x,b,Q)}$ is the Sudakov factor. The function $s(x,b,Q)$ up to next-to-leading order has been given in Ref. \cite{Li1995}, which is  
%首先定义两个核心对数变量：

\begin{widetext}
\begin{equation}
	\begin{aligned}
		s(x,b,Q) &= \frac{A^{(1)}}{2\beta_1}\hat{q}\ln\left(\frac{\hat{q}}{\hat{b}}\right) - \frac{A^{(1)}}{2\beta_1}(\hat{q}-\hat{b}) + \frac{A^{(2)}}{4\beta_1^2}\left(\frac{\hat{q}}{\hat{b}}-1\right) - \left[\frac{A^{(2)}}{4\beta_1^2}-\frac{A^{(1)}}{4\beta_1}\ln\left(\frac{e^{2\gamma_E-1}}{2}\right)\right]\ln\left(\frac{\hat{q}}{\hat{b}}\right) \\
		&\quad + \frac{A^{(1)}\beta_2}{4\beta_1^3}\hat{q}\left[\frac{\ln(2\hat{q})+1}{\hat{q}} - \frac{\ln(2\hat{b})+1}{\hat{b}}\right] + \frac{A^{(1)}\beta_2}{8\beta_1^3}\left[\ln^2(2\hat{q}) - \ln^2(2\hat{b})\right] \\
		&\quad + \frac{A^{(1)}\beta_2}{8\beta_1^3}\ln\left(\frac{e^{2\gamma_E-1}}{2}\right)\left[\frac{\ln(2\hat{q})+1}{\hat{q}} - \frac{\ln(2\hat{b})+1}{\hat{b}}\right] - \frac{A^{(2)}\beta_2}{16\beta_1^4}\left[\frac{2\ln(2\hat{q})+3}{\hat{q}} - \frac{2\ln(2\hat{b})+3}{\hat{b}}\right] \\
		&\quad - \frac{A^{(1)}\beta_2}{16\beta_1^4}\frac{\hat{q}-\hat{b}}{\hat{b}^2}[2\ln(2\hat{b})+1] + \frac{A^{(2)}\beta_2^2}{432\beta_1^6}\frac{\hat{q}-\hat{b}}{\hat{b}^3}[9\ln^2(2\hat{b})+6\ln(2\hat{b})+2] \\
		&\quad + \frac{A^{(2)}\beta_2^2}{1728\beta_1^6}\left[\frac{18\ln^2(2\hat{q})+30\ln(2\hat{q})+19}{\hat{q}^2} - \frac{18\ln^2(2\hat{b})+30\ln(2\hat{b})+19}{\hat{b}^2}\right],
	\end{aligned}
\end{equation}
\end{widetext}
where
\begin{equation}
	\hat{q} = \ln\left(\frac{x Q_m}{\sqrt{2}\Lambda}\right), \quad \hat{b} = \ln\left(\frac{1}{b\Lambda}\right),
\end{equation}
\begin{equation}
	\begin{aligned}
		\beta_1 &= \frac{33 - 2n_f}{12}, \quad \beta_2 = \frac{153 - 19n_f}{24}, \quad A^{(1)} = \frac{4}{3}, \\
		A^{(2)} &= \frac{67}{9} - \frac{\pi^2}{3} - \frac{10}{27}n_f + \frac{8}{3}\beta_1 \ln \left( \frac{e^{\gamma_E}}{2} \right),
	\end{aligned}
\end{equation}
with $\gamma_E$ being the Euler constant, and $\gamma_E=0.57721$.

The threshold factor $S_t(x)$ can be parameterized as \cite{Li2002,KLS-2001}
\begin{equation}
	S_t(x) = \frac{2^{1+2c} \Gamma(3/2+c)}{\sqrt{\pi} \Gamma(1+c)} [x(1-x)]^c,
\end{equation}
where $c$ is determined to be 0.3.

\section{WAVE FUNCTION} \label{Appendix B}
	
%\begin{equation}
%	\begin{aligned}
%		\langle \rho / \omega(P, \epsilon_V^T) | \bar{d}_\alpha(z) u_\beta(0) | 0 \rangle &= \frac{1}%{\sqrt{2N_c}} \int_0^1 dx e^{ixP \cdot z} \Big[ M_V [\slashed{\epsilon}_V^{*T}] \phi_V^v(x) \\
%		&+ [\slashed{\epsilon}_V^{*T} \slashed{P}] \phi_V^T(x) - \frac{M_V}{P \cdot n_+} i %\epsilon_{\mu\nu\rho\sigma} [\gamma^5 \gamma^\mu] \epsilon_V^{T\nu} P^\rho n_+^\sigma \phi_V^a(x) \Big].
%	\end{aligned}
%\end{equation}

The distributin amplitudes for the vector meson $\rho$ can be expressed in terms of the Gegenbauer polynomials as \cite{Ball-1998} 
\begin{gather}
	\phi_{\rho}^{T}(x) = \frac{3 f_{\rho}^{T}}{\sqrt{6}} x(1 - x) [1 + 0.2 C_{2}^{3/2}(t)], \\
\begin{split}
	\phi_{\rho}^{v}(x) &= \frac{f_{\rho}}{2 \sqrt{6}} \left[ \frac{3}{4}(1 + t^{2}) + 0.24(3t^{2} - 1)\right. \\
&\quad+ 0.12(3 - 30t^{2} + 35t^{4}) \left.\right],
\end{split} \\
	\phi_{\rho}^{a}(x) = \frac{3 f_{\rho}}{4 \sqrt{6}} t [1 + 0.93(10x^{2} - 10x + 1)],
\end{gather}
and
\begin{align}
	\phi_V^T(x) &= \frac{f_V^T}{2\sqrt{2N_c}} \phi_\perp, \\
	\phi_V^v(x) &= \frac{f_V}{2\sqrt{2N_c}} g_\perp^{(v)}, \\
	\phi_V^a(x) &= \frac{f_V}{8\sqrt{2N_c}} \frac{d}{dx} g_\perp^{(a)},
\end{align}
with the normalization condition
\begin{equation}
	\int_{0}^{1} dx \phi_{i}(x) = 1,
\end{equation}
where $\phi_{i}=\{\phi_\perp,g_\perp^{(v)},g_\perp^{(a)}\}$, and
\begin{equation}
	C_{2}^{3/2}(t) = \frac{3}{2}(5t^2 - 1).
\end{equation}
For the distribution amplitudes of $\omega$, we use the same functions as that of $\rho$ meson because of the similarity of these two vector mesons.
	
	% -------------------------------------------------------------------
	% 参考文献
	% -------------------------------------------------------------------
	% 推荐使用 BibTeX 管理参考文献
	% 创建一个名为 references.bib 的文件，并在其中添加 BibTeX 条目
	
	% \bibliography{references} 
	
	% 如果你暂时不想用 .bib 文件，也可以手动写 (不推荐)：

\end{document}